\documentclass[a4paper,onecolumn,11pt,accepted=2026-07-28]{quantumarticle}
\pdfoutput=1
\usepackage[utf8]{inputenc}
\usepackage[english]{babel}
\usepackage[T1]{fontenc}
\usepackage{amsmath}
\usepackage{hyperref}

\usepackage{tikz}
\usepackage{lipsum}

\usepackage{graphicx}
\usepackage{dcolumn}
\usepackage{bm}
\usepackage{braket}
\usepackage{float}
\usepackage[dvipsnames]{xcolor}
\usepackage{qcircuit}
\usepackage{amssymb}
\usepackage{multirow}

\def\[#1\]{\begin{align}#1\end{align}}

\def\f{\int_{-\infty}^\infty}

\def\ex[#1]{\exp\left(#1\right)}
\def\ps{\ket{\psi}}
\def\nn{\nonumber\\}
\def\b[#1]{#1}

\begin{document}
\title{Quantum phase estimation with optimal confidence interval using three control qubits}
\author{Kaur Kristjuhan}
\affiliation{Molecular Quantum Solutions ApS, Blegdamsvej 17, 2100 Copenhagen, Denmark}
\affiliation{School of Mathematical and Physical Sciences, Macquarie University, Sydney, NSW 2109, Australia}
\affiliation{Sydney Quantum Academy, Sydney, NSW, Australia}
\email{kaur@mqs.dk}
\author{Dominic W. Berry}
\affiliation{School of Mathematical and Physical Sciences, Macquarie University, Sydney, NSW 2109, Australia}
\maketitle
\begin{abstract}
Quantum phase estimation is an important routine in many quantum algorithms, particularly for estimating the ground state energy in quantum chemistry simulations. This estimation involves applying powers of a unitary to the ground state, controlled by an auxiliary state prepared on a control register. In many applications the goal is to provide a confidence interval for the phase estimate, and optimal performance is provided by a discrete prolate spheroidal sequence. We show how to prepare the corresponding state in a far more efficient way than prior work. We find that a matrix product state representation with a bond dimension of 4 is sufficient to give a highly accurate approximation for all dimensions tested, up to $2^{24}$. This matrix product state can be efficiently prepared using a sequence of simple three-qubit operations. When the dimension is a power of 2, the phase estimation can be performed with only three qubits for the control register, making it suitable for early-generation fault-tolerant quantum computers with a limited number of logical qubits. 
\end{abstract}

\section{Introduction}
Quantum phase estimation (QPE) is a widely used subroutine in quantum computing and is indispensable for many important applications such as factorisation \cite{shor1999polynomial} and ground state energy estimation in quantum chemistry \cite{aspuru2005simulated, mcardle2020quantum, cao2019quantum, bauer2020quantum, motta2022emerging, dalzell2023quantum}. The purpose of QPE is to estimate the phase $\phi$ of an eigenvalue $e^{i\phi}$ of a unitary operator $U$. The algorithm for QPE \cite{nielsen2010quantum} invokes the application of controlled powers of $U$ on a quantum state many times.

The estimate can be improved to arbitrary precision by using larger powers of $U$.
In the case of factoring, these powers may be performed with similar complexity to $U$, so it is convenient to use the simple textbook form of QPE with large powers of $U$.
In contrast, for ground state energy estimation, $U$ is an operator that encodes the Hamiltonian describing the physical system of interest.
In that context, higher powers of $U$ correspond to Hamiltonian simulations with longer evolution time, or more applications of the qubitised operator \cite{BerryNPJ18,Poulin18}.
This means that the complexity increases linearly in the power of $U$, and these powers constitute the majority of the cost of the algorithm.

This motivates us to find ways to modify the QPE algorithm to maximise the accuracy of the phase estimate with minimal applications of $U$.
The first step of the QPE algorithm, as described in most textbooks \cite{nielsen2010quantum}, is to prepare an equal superposition state on a control register. Replacement of this step with preparation of a different state is one way to improve the accuracy of QPE.

\b[A closely related modification is used in classical signal processing. Assigning equal weight to every sample in a dataset is called a rectangular window, and it is well known to cause problems in spectral analysis \cite{harris1978use}.
Uniform weighting causes the discrete Fourier transform to spread energy across neighbouring frequency bins, rather than concentrating it at the true frequency. The solution is to replace the rectangular window with a smoother window function that suppresses this spectral leakage.
Precision metrology faces the same issue, and gravitational wave observatories such as LIGO taper the edges of their time series data before Fourier analysis for this reason \cite{abbott2017gw170817,Usman_2016}.
Quantum phase estimation relies on an analogous Fourier transform between the control register and the estimated phase, and so encounters the same leakage issue. 
We can adapt these classical tapering techniques to the quantum setting by applying a smooth amplitude profile across the control register. This acts as a quantum window function, which concentrates the measured phase distribution around the true eigenphase.
]

If the goal is to perform measurements with minimum mean-square error, then the control register can be in a state with amplitudes according to a sine function, which is easily prepared \cite{luis1996optimum,Childs_2009,babbush2018encoding,najafi2023optimum}.
\b[
This provides the optimal average case accuracy for QPE, but does not provide any guarantees of the estimate being within a desired error of the true answer.
One example where this is especially relevant is ground state energy estimation, in the case where we are unable to precisely prepare the ground state.
In such a situation, the phase measurement may erroneously probe the energy of an excited state instead.
To isolate the ground state from excited states, we can take multiple samples and select the minimum value. 
However, this strategy requires a careful balance: under-sampling risks missing the ground state entirely, while over-sampling increases the likelihood of underestimating the true energy.]

\b[The alternative is to optimise for the confidence interval of the phase estimate instead.
This minimises the probability of obtaining a result with large error \cite{berry2024rapid}.] 
The use of a discrete prolate spheroidal sequence (DPSS) state in the control register has been shown to minimise the width of the confidence interval \cite{imai2009fourier} \b[and also enables QPE to be used as a viable technique for filtering \cite{Sakuma_2026}].

\b[The DPSS state shares a direct physical analogue with optimal apodisation in classical optics \cite{slepian65}. 
Just as high-precision optical systems use prolate spheroidal transmission masks to smoothly taper a beam's intensity profile and maximise the concentration of focused light by suppressing diffraction rings \cite{thomas2008testing, Xie:10}, the DPSS state maximally concentrates the quantum measurement probability around the true phase.]

Due to the difficulty of calculating the DPSS \cite{slepian1978prolate}, some works have opted to use the Kaiser window state instead \cite{kaiser1980use}, which approximates the DPSS state with good accuracy and has asymptotic behaviour that can be analytically described \cite{berry2024analyzing, berry2024rapid}.
Both the DPSS state and Kaiser window state are considerably more complicated than the sine window from Ref.~\cite{babbush2018encoding,najafi2023optimum}, so a new state preparation scheme is needed.

For preparing the DPSS state, variations of Grover-Rudolph state preparation \cite{grover2002creating, lowTradingGatesDirty2024a} or state preparation based on the quantum singular value transformation are commonly proposed methods \cite{greenaway2024case, o2024quantum, mcardle2022quantum}. Reference~\cite{Patel_2026} also describes a specialised state preparation procedure for the DPSS state\b[, which involves discarding low amplitude components of the DPSS window in frequency space to enable an approximate, but less costly preparation]. For all of these approaches, the number of non-Clifford gates scales polynomially in the dimension (so exponentially in the number of qubits).
Moreover, all qubits for the state are prepared at once, and many more additional qubits may be used in schemes based on quantum read-only memory (QROM) \cite{lowTradingGatesDirty2024a, berry2024rapid}.
These features are particularly undesirable for early generations of fault-tolerant quantum computers, where qubit counts are limited and gate operations are slow.

We propose a more resource-efficient state preparation procedure to prepare a DPSS state of any dimension based on a matrix product state (MPS) representation. The implementation cost of the state preparation grows linearly with the number of qubits in the control register and enables the user to perform the minimal number of controlled applications of $U$ to achieve a target confidence interval width. Furthermore, if the dimension of the DPSS state is a power of two, we are able to exploit the structure of the MPS preparation circuit to sequentially measure and recycle the control register qubits used during phase estimation so that only three ancilla qubits are required for implementation.

The remainder of this paper is structured as follows. Section \ref{background} explains the necessary background theory of QPE, DPSS states, and MPS approximations. In Section \ref{results} we provide numerical evidence that using an MPS approximation with bond dimension 4 is suitable for preparing a DPSS state and provides the optimal confidence interval at any desired precision for confidence levels used in practical applications. Finally, in Section \ref{implementation} we propose an efficient gate implementation for preparing an MPS with real-valued amplitudes, provide explicit quantum circuits comprised of single qubit rotations and Clifford gates, and demonstrate reduced implementation cost compared to prior work.

\section{Quantum phase estimation with discrete prolate spheroidal sequence states}\label{background}

\subsection{Overview of quantum phase estimation}\label{qpe}
In this section, we will provide an overview of the QPE algorithm. This mainly reviews prior work, but we also provide some more explicit explanations of how a DPSS state can be used to enhance the performance of the algorithm and discuss alternative implementations where the dimension of the state is not necessarily a power of two.

The quantum phase estimation algorithm begins with the premise that we have a system register where we have already prepared an eigenstate $\ket{\phi}$ of an operator $U$. The task is to estimate the phase of the eigenvalue $\phi$ corresponding to that state, meaning that $U\ket{\phi}=e^{i\phi}\ket{\phi}$. The first step of the algorithm is to prepare a $D$-dimensional state $\ps=\sum_{\b[k]=0}^{D-1}\gamma_k\ket{k}$ on a control register comprising $n$ qubits. In the textbook version of QPE \cite{nielsen2010quantum}, the amplitudes are equal and $D$ is a power of two.

The next step is to apply the operator $U^k$ to the system register, where $k$ is controlled on each value of $k$ in the control state. The resulting state is
\[
\ps\ket{\phi}\mapsto\sum_{k=0}^{D-1}\gamma_k\ket{k}U^k\ket{\phi}=\sum_{k=0}^{D-1}\gamma_ke^{i\phi k}\ket{k}\ket{\phi}.
\]
Instead of explicitly performing $U^k$ controlled on each value of $k$, the most common implementation is to apply powers of $U$ that are powers of two, each controlled on just one of the qubits in the control register. This saves considerable resources if $U^k$ can be implemented with lower cost than $k$ applications of $U$. This is true in the context of factoring \cite{shor1999polynomial}, but not for ground state energy estimation problems, where the cost of $U^k$ is linear in $k$ at best, as $k$ corresponds to the simulation time in Hamiltonian simulation algorithms \cite{low2017optimal}.

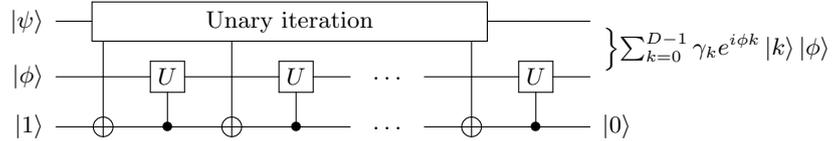
\begin{figure}[tbh]
    \centering
~\\~\\~\\
\begin{tikzpicture}[remember picture,overlay]
  \node (circuit) {
    \Qcircuit @C=1.5em @R=1em {
      \lstick{\ps}& \ctrl{2} &\qw& \ctrl{2} & \gate{U}&\qw&\dots&&\ctrl{2}&\qw&\qw \\
      \lstick{\ket{\phi}}&\qw& \gate{U} & \qw & \gate{U}&\qw&\dots&&\qw&\gate{U}&\qw\\
  \lstick{\ket{1}}& \targ&\ctrl{-1}& \targ      & \ctrl{-1}&\qw&\dots&&\targ&\ctrl{-1}&\rstick{\ket{0}}\qw\\
    }
  };
  \node at (6.0,0.33) {$\left\}\sum_{k=0}^{D-1}\gamma_ke^{i\phi k}\ket{k}\ket{\phi}\right.$};
  \node[draw=black, fill=white, 
      minimum width=16em, minimum height=1.6em, text centered] 
      at ([yshift=-1em, xshift=-1.35em]circuit.north) {Unary iteration};
\end{tikzpicture}
\\~\\~\\
    \caption{Variant of the first steps of the QPE algorithm with $D-1$ applications of controlled unitary operations, where $D$ may be any positive integer.}
    \label{unary}
\end{figure}

One way to reduce the cost of this step is to opt for a smaller value of $D$, at the expense of the accuracy of the result. To provide the most flexibility in this choice, it would ideally be possible to select any value of $D$ which need not be a power of two. This entails modifying $\ps$ accordingly and then performing a unary iteration \cite{babbush2018encoding} procedure on the control register \cite{LeePRXQ21}.
During each step of the unary iteration, the operator $U$ would be applied to the system register, controlled on an ancilla qubit, which is initialised in the $\ket{1}$ state, as shown in Figure \ref{unary}. The unary iteration then controls the flip of the ancilla, meaning that if the control register is in the state $\ket{k}$, then the ancilla is flipped and $U$ will have been applied to the system register $k$ times. This procedure achieves the desired effect of applying $U^k$ to the system register, using a total of $D-1$ applications of controlled $U$, where $D$ can be any integer.
It is also possible to increase the precision by controlling between $U$ and $U^\dagger$ \cite{babbush2018encoding}.
For $U$ a qubitised operator \cite{BerryNPJ18,Poulin18}, the unary iteration can instead be used to control the reflection, and the extra ancilla qubit is not needed \cite{LeePRXQ21}.

The final step of QPE is to apply an inverse quantum Fourier transform (QFT) to the control register and perform a measurement in the computational basis. The state before measurement is
\[
\sum_{k=0}^{D-1}\gamma_ke^{i\phi k}\ket{k}\ket{\phi}\overset{\text{QFT}^{-1}}{\mapsto}\frac{1}{\sqrt{N}}\sum_{k=0}^{D-1}\sum_{l=0}^{N-1}\gamma_k
e^{ik(\phi-2\pi l/N)}
\ket{l}\ket{\phi}=\sum_{l=0}^{N-1}\Gamma\!\left(\phi-\frac{2\pi l}{N}\right)\ket{l}\ket{\phi},
\]
where $N=2^n\ge D$ and we have defined
\[
\Gamma(x)\equiv\frac{1}{\sqrt{N}}\sum_{k=0}^{D-1}\gamma_k e^{ikx}.
\]
The value of $N$ is chosen as a power of 2, because that enables the inverse QFT to be performed efficiently.
Based on the measurement, we obtain an estimate of $\phi$, which we denote $\hat{\phi}$. A commonly used procedure is to set the estimate as $\hat{\phi}=2\pi l/N$ for a measurement outcome $l$, which occurs with probability $|\Gamma(\phi-2\pi l/N)|^2=|\Gamma(\phi-\hat{\phi})|^2$.

A disadvantage of doing so is that it restricts our phase estimates to a discrete set of outcomes. This induces an error in our estimate that depends on how close the true value of $\phi$ is to the values in this set. This feature is undesirable because the true value of $\phi$ is unknown and in the worst case scenario where $\phi$ is precisely between two consecutive multiples of $2\pi/N$, the phase estimation would always yield an error of at least $\pi/N$, resulting in larger average error than for other values of $\phi$.
This method of performing the phase estimation complicates the error analysis, because the performance of QPE averaged over all possible values of $\phi$ can be different from the performance of QPE for a particular unknown $\phi$.

A known solution to this is to add a randomly chosen, known phase $\tilde{\phi}$ to $\phi$ within the QPE circuit, such that $-\pi/N<\tilde{\phi}\leq\pi/N$ \cite{berry2012optimal}. We explain this approach below to clarify how it can be applied in practice. The probability density function of the continuous variable $\tilde{\phi}$ is a uniform distribution. 
We will use $p(\cdot)$ to denote probability densities of continuous variables and $\text{Pr}(\cdot)$ to denote probabilities of discrete outcomes, with the argument specifying the variable under consideration in each instance.
The probability density of $\tilde{\phi}$ is
\[\label{pu}
p(\tilde{\phi})=\left\{\begin{array}{ll}
\frac{N}{2\pi}\text{ for }-\pi/N<\tilde{\phi}\leq\pi/N\\
0\text{ elsewhere}
\end{array}.\right.
\]
Implementing this phase shift can be achieved by inserting a phase \b[shift] gate on each qubit in the control register after the state preparation step, where the applied phase is $2^j\tilde{\phi}$ with $j$ an integer incremented by 1 for each qubit in the control register, starting from $j=1$ for the least significant qubit and ending with $j=n$ for the most significant qubit. The range we have defined for $\tilde{\phi}$ ensures that $-\pi<2^j\tilde{\phi}\leq\pi$ for all values of $j$. After doing this, the state on the control register is $\sum_{k=0}^{D-1}\gamma_k e^{i(\phi+\tilde{\phi})k}\ket{k}\ket{\phi}$, which becomes $\sum_{l=0}^{N-1}\Gamma\!\left(\phi+\tilde{\phi}-\frac{2\pi l}{N}\right)\ket{l}\ket{\phi}$ after applying the inverse quantum Fourier transform.

Proceeding with the measurement, we obtain a measurement outcome $l$ with the conditional probability
\[
\text{Pr}(l|\phi,\tilde{\phi})=\left|\Gamma\!\left(\phi+\tilde{\phi}-\frac{2\pi l}{N}\right)\right|^2 .
\]
We can define a continuous variable $y$ to express the associated conditional probability density function of the discrete variable $l$:
\[
p(y|\phi,\tilde{\phi})\equiv\sum_{l=0}^{N-1}\text{Pr}(l|\phi,\tilde{\phi})\delta(y-l) .
\]
We can now define the estimate of $\phi$ as a continuous variable $\hat{\phi}=2\pi y/N - \tilde{\phi}$, where the subtraction of $\tilde{\phi}$ is accounting for the adjustment made in the circuit. Inserting $y=(\hat{\phi}+\tilde{\phi})N/2\pi$ into the previous expression yields the conditional probability density of $\hat{\phi}$:
\[
p(\hat{\phi}|\phi,\tilde{\phi})=\b[\frac{N}{2\pi}]\sum_{l=0}^{N-1}\text{Pr}(l|\phi,\tilde{\phi})\,\delta\!\left(\frac{(\hat{\phi}+\tilde{\phi})N}{2\pi}-l\right),
\]
\b[which has been multiplied by a factor of $N/2\pi$ to ensure normalisation of the probability density upon integrating over the new variable, as $dy=(N/2\pi)d\hat{\phi}$]. The marginal probability density of $\hat{\phi}$, obtained by averaging all possible values of $\tilde{\phi}$, is
\[\label{margprob}
p(\hat{\phi}|\phi)&=\f p(\hat{\phi}|\phi,\tilde{\phi})p(\tilde{\phi})\, d\tilde{\phi}=\b[\frac{N}{2\pi}]\sum_{l=0}^{N-1}\f \text{Pr}(l|\phi,\tilde{\phi})\,\delta\!\left(\frac{(\hat{\phi}+\tilde{\phi})N}{2\pi}-l\right)p(\tilde{\phi})\, d\tilde{\phi}\nn
&=\left.\sum_{l=0}^{N-1} \text{Pr}(l|\phi,\tilde{\phi})p(\tilde{\phi})\right|_{\tilde{\phi}=2\pi l/N-\hat{\phi}}=\left.|\Gamma(\phi-\hat{\phi})|^2\sum_{l=0}^{N-1} p(\tilde{\phi})\right|_{\tilde{\phi}=2\pi l/N-\hat{\phi}},
\]
where we have used the property of the Dirac delta function that if $g(x)=0$ only for $x=x_0$, then
\[
\f f(x)\delta(g(x))dx=\frac{f(x_0)}{g'(x_0)}.
\]
Regardless of the value of $\hat{\phi}$, Eq.~\eqref{pu} ensures that the expression $p(\tilde{\phi})|_{\tilde{\phi}=2\pi l/N-\hat{\phi}}$ has the value $N/2\pi$ for exactly one value of $l$ and is zero elsewhere, so 
\[
\left.\sum_{l=0}^{N-1} p(\tilde{\phi})\right|_{\tilde{\phi}=2\pi l/N-\hat{\phi}}=\frac{N}{2\pi},
\]
and Eq.~\eqref{margprob} reduces to
\[
p(\hat{\phi}|\phi)=\b[\frac{N}{2\pi}]|\Gamma(\phi-\hat{\phi})|^2.
\]
The probability density of the error $\Delta\phi\equiv\phi-\hat{\phi}$ is therefore
\[
p(\Delta\phi)=\b[\frac{N}{2\pi}]|\Gamma(\Delta\phi)|^2,
\]
which is independent of the value of $\phi$, as desired.
The confidence level $1-\delta$ with which the true value $\phi$ lies within a confidence interval of width $2d$ centred around the estimate $\hat{\phi}$ can be calculated as
\[
1-\delta=\text{Pr}[-d\leq\Delta\phi\leq d]=\int_{-d}^{d}p(\Delta\phi)d\Delta\phi=\b[\frac{N}{2\pi}]\int_{-d}^d|\Gamma(\Delta\phi)|^2d\Delta\phi.
\]
\b[For this analysis we require $0\leq d\leq\pi$, because the confidence interval is defined on the circle]. Determining the coefficients $\gamma_k$ that maximise the value of this expression has been studied under the name of the spectral concentration problem, where the objective is to find a bandlimited sequence with maximal concentration \cite{simons2010slepian}. For the discrete case, the sequence $\{\gamma_k\}$ which solves this problem is the first DPSS \cite{slepian1978prolate}. Each DPSS can be defined as a solution to the eigenvalue problem $S(d)\ps=\lambda(d)\ps$, where
\[
S(d)\equiv\frac{d}{\pi}\sum_{j=0}^{D-1}\sum_{k=0}^{D-1}\frac{\sin(d(j-k))}{d(j-k)}\ket{j}\bra{k}\label{sop}.
\]
The first DPSS corresponds to the solution with the largest eigenvalue, which is equal to the confidence level $1-\delta$, and throughout the paper we refer to the DPSS state as the corresponding eigenstate. The DPSS state therefore provides the largest possible confidence level for a given confidence interval width, or conversely, achieves a desired confidence level at the tightest possible confidence interval. The next sections focus on how to efficiently prepare the DPSS state on the control register using a matrix product state approximation.

\subsection{Matrix product states}
A matrix product state $\ket{\psi_\text{MPS}}$ for $n$ subsystems describes a state where a set of matrices $W^{[k]}$ are associated with each subsystem $k$ and the wavefunction of the total system is expressed using a product of these matrices:
\[
\ket{\psi_\text{MPS}}=\sum_{s_1,\dots,s_n}\text{Tr}\left(W^{[1]}_{s_1}W^{[2]}_{s_2}\dots W^{[n]}_{s_n}\right)\ket{s_1\dots s_n},
\]
where each $W^{[k]}_{s_k}$ is a matrix prescribed to subsystem $k$ in the state $\ket{s_k}$. The number of columns in one matrix must correspond to the number of rows in the next matrix so that matrix multiplication is properly defined. This number shared between two neighbouring matrices is called the bond dimension $\chi$.
The indices that span the rows and columns of these matrices are called virtual indices $\alpha_k$, to distinguish them from physical indices $s_k$, which describe the subsystems. An MPS with bond dimension $\chi$ is an MPS where $\chi$ is the maximum bond dimension among all neighbouring pairs of matrices. If the subsystems are qubits, then any quantum state can be expressed exactly as an MPS with bond dimension $2^{\lfloor n/2\rfloor}$ \cite{vidal2003efficient,schollwock2011density}.

If we restrict $\chi$ to a fixed value less than $2^{\lfloor n/2\rfloor}$, then only a subset of $n$-qubit states can be represented. However, an arbitrary state can be approximated by choosing the closest state from this subset. The quality of the approximation depends on both the target state and the bond dimension.
The state that provides the best approximation from this subset can be found with an algorithm based on singular value decomposition (SVD) \cite{schollwock2011density}. This can be applied in different ways, producing different representations of the same state \cite{schollwock2011density, vidal2003efficient}. We are most interested in representations where the matrices $W^{[k]}_{s_k}$ are unitaries or isometries that can be implemented directly in a quantum circuit.

To implement each $W^{[k]}_{s_k}$ as an operator, these matrices are first calculated classically. Then, a procedure for unitary synthesis \cite{dawson2005solovay, hao2025reducing, amy2013meet, gheorghiu2022t, paradis2024synthetiq, weiden2024high} or isometry synthesis \cite{berry2024rapid, lowTradingGatesDirty2024a, itenQuantumCircuitsIsometries2016, szaszNumericalCircuitSynthesis2023} is used to generate a quantum circuit that implements these matrices. Increasing the bond dimension increases the amount of classical data that needs to be incorporated into the circuit, thereby incurring a $T$-gate cost scaling as $\mathcal{O}(\chi^{3/2})$ \cite{fomichev2024initial}. The number of qubits that each operator acts on also scales logarithmically with bond dimension, further increasing the size and complexity of the circuit as bond dimension is increased. For these reasons, it is of practical interest to use MPS approximations of states that have minimal bond dimension while providing sufficient accuracy.

\subsection{Preparation of DPSS states}\label{prep}
To prepare a quantum state $\ps$ that spans $n$ qubits as an MPS, we can follow the procedure described in Ref.~\cite{schollwock2011density}, which involves \b[classically computing ]successive applications of reshaping and singular value decomposition (SVD) of the state.
\b[In this section we explain this procedure in detail, showing how the state preparation quantum circuit is constructed from the target state.
A small-scale example is provided in Appendix \ref{appc}.]

\begin{figure}[tbh]
    \centering
    \[
\Qcircuit @R=1em @C=.7em {
\lstick{\ket{0}}&{/}\qw&\multigate{1}{M^{[k]}}&{/}\qw&\rstick{\alpha_k}\qw&\rstick{\quad(\textbf{a})}\\
\lstick{\alpha_{k-1}}&{/}\qw&\ghost{M^{[k]}}&\qw&\rstick{s_k}\qw
}
\qquad\qquad\qquad\qquad\;\;
\Qcircuit @R=1em @C=.7em {
\lstick{\alpha_{k-1}}&{/}\qw&\multigate{1}{V^{[k]}}&\qw&\rstick{s_k}\qw&\rstick{\quad(\textbf{b})}\\
\lstick{\ket{0}}&{/}\qw&\ghost{V^{[k]}}&{/}\qw&\rstick{\alpha_k}\qw
}
    \nonumber
\]
\[
\Qcircuit @R=1em @C=.7em {
&&&&&&&&&&&\multigate{2}{M^{[k+1]}}&\qw&\rstick{(\textbf{c})}\\
&&&\push{\ket{0}}&&\qw&\multigate{2}{M^{[k]}}&\qw&\push{\alpha_k}&&\qw&\ghost{M^{[k+1]}}&\qw\\
&\multigate{2}{M^{[k-1]}}&\qw&\push{\alpha_{k-1}}&&\qw&\ghost{M^{[k]}}&\qw&\push{\alpha_k}&&\qw&\ghost{M^{[k+1]}}&\qw\\
&\ghost{M^{[k-1]}}&\qw&\push{\alpha_{k-1}}&&\qw&\ghost{M^{[k]}}&\rstick{s_k}\qw\\
&\ghost{M^{[k-1]}}&&&&&&
}
\nonumber
\]
    \caption{Diagram (a) illustrates an operator $M^{[k]}$, which can be constructed by reordering the rows and columns of $V^{[k]}$, shown in diagram (b). The bottommost qubit is the most significant qubit. The slashes on the lines indicate that the number of qubits needed varies according to the dimension needed to span the corresponding index. Diagram (c) shows how an operator $M^{[k]}$ connects to neighbouring operators in an MPS. For $\chi=4$, most operators in the MPS will have two qubits indexing $\alpha_{k-1}$ and $\alpha_k$, and one qubit in the state $\ket{0}$ needed to match the dimension on either side.}
    \label{vandw}
\end{figure}
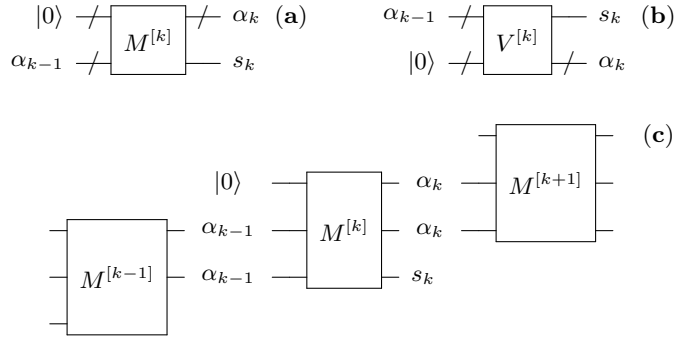

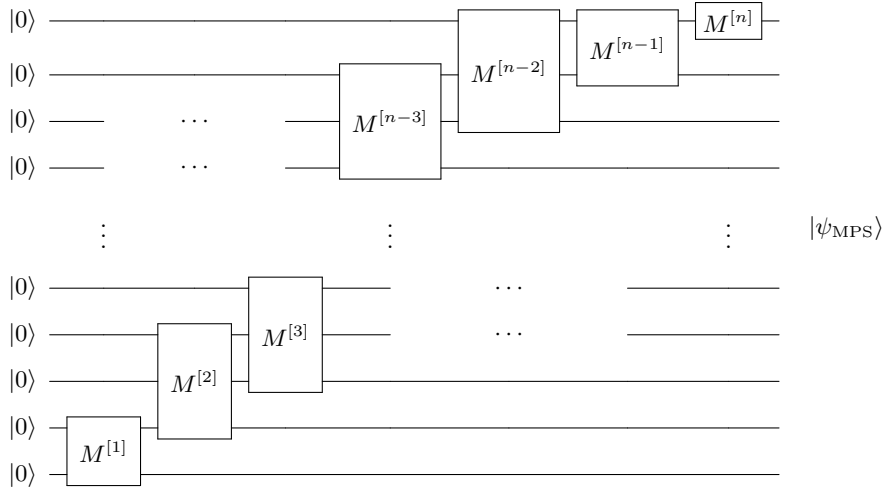
\begin{figure}[tbh]
    \centering
    \[
\Qcircuit @R=1em @C=.7em {
\lstick{\ket{0}}&\qw&\qw&\qw&\qw&\multigate{2}{M^{[n-2]}}&\multigate{1}{M^{[n-1]}}&\gate{M^{[n]}}&\qw\\
\lstick{\ket{0}}&\qw&\qw&\qw&\multigate{2}{M^{[n-3]}}&\ghost{M^{[n-2]}}&\ghost{M^{[n-1]}}&\qw&\qw\\
\lstick{\ket{0}}&\qw&\cdots&&\ghost{M^{[n-3]}}&\ghost{M^{[n-2]}}&\qw&\qw&\qw\\
\lstick{\ket{0}}&\qw&\cdots&&\ghost{M^{[n-3]}}&\qw&\qw&\qw&\qw\\
&&&&&&&&&\\
&\vdots&&&\vdots&&&\vdots&&\rstick{\ket{\psi_\text{MPS}}}\\
&&&&&&&&&\\
\lstick{\ket{0}}&\qw&\qw&\multigate{2}{M^{[3]}}&\qw&\cdots&&\qw&\qw&\\
\lstick{\ket{0}}&\qw&\multigate{2}{M^{[2]}}&\ghost{M^{[3]}}&\qw&\cdots&&\qw&\qw\\
\lstick{\ket{0}}&\qw&\ghost{M^{[2]}}&\ghost{M^{[3]}}&\qw&\qw&\qw&\qw&\qw\\
\lstick{\ket{0}}&\multigate{1}{M^{[1]}}&\ghost{M^{[2]}}&\qw&\qw&\qw&\qw&\qw&\qw\\
\lstick{\ket{0}}&\ghost{M^{[1]}}&\qw&\qw&\qw&\qw&\qw&\qw&\qw
}
    \nonumber
    \]
    \caption{Quantum circuit for preparing an MPS with $\chi=4$. All operators between $M^{[2]}$ and $M^{[n-2]}$ act on three consecutive qubits. The bottommost qubit is the most significant qubit.}
    \label{circuit}
\end{figure}

We start with the amplitude vector $\psi_{s_1\dots s_n}$of the target state
\[
\ps=\sum_{s_1\dots s_n}\psi_{s_1\dots s_n}\ket{s_1\dots s_n},
\]
and reshape it into a rectangular matrix of dimension $2^{n-1}\times 2$, which we denote by $\psi_{s_1\dots s_{n-1},s_n}$ (the comma separates the row index from the column index).
For this state, $s_1$ refers to the most significant qubit and $s_n$ refers to the least significant qubit.
Performing an SVD on this matrix yields $\psi_{s_1\dots s_{n-1},s_n}=\sum_{\alpha_{n-1}}\psi_{s_1\dots s_{n-1},\alpha_{n-1}} V^{\dagger[n]}_{\alpha_{n-1},s_n}$, where $V^{\dagger[n]}$ has orthonormal rows containing the right singular vectors and $\psi_{s_1\dots s_{n-1},\alpha_{n-1}}$ is the remaining part of the SVD. We then repeat this process, each time reshaping the remaining part so that one physical index becomes a column index
\[
\psi_{s_1\dots s_k,\alpha_k}\rightarrow\psi_{s_1\dots s_{k-1},s_k\alpha_k}=\sum_{\alpha_{k-1}}\psi_{s_1\dots s_{k-1},\alpha_{k-1}}V^{\dagger[k]}_{\alpha_{k-1},s_k\alpha_k}.
\]
After performing this procedure for all sites, we obtain an MPS in the right canonical form
\[\label{reshape_and_svd}
\psi_{s_1\dots s_n}=\sum_{\alpha_1\dots\alpha_{n-1}}V^{\dagger[1]}_{s_1\alpha_1}V^{\dagger[2]}_{\alpha_1,s_2\alpha_2}\dots V^{\dagger[n-1]}_{\alpha_{n-2},s_{n-1}\alpha_{n-1}}V^{\dagger[n]}_{\alpha_{n-1},s_n},
\]
where $V^\dagger_{s_1\alpha_1}$ is a row vector instead of a matrix. Clearly, this is a matrix product state with $W_{s_k}^{[k]}=V^{\dagger[k]}_{\alpha_{k-1},s_k\alpha_k}$. For real-valued amplitudes, we can make use of the trace property $\text{Tr}(X)=\text{Tr}(X^\dagger)^*$ to rewrite the amplitudes as
\[
\psi_{s_1\dots s_n}&=\psi_{s_1\dots s_n}^*=\text{Tr}\left(W^{[1]}_{s_1}W^{[2]}_{s_2}\dots W^{[n]}_{s_n}\right)^*=\text{Tr}\left(W^{\dagger[n]}_{s_n}\dots W^{\dagger[2]}_{s_2}\dots W^{\dagger[1]}_{s_1}\right)\nonumber\\&=\sum_{\alpha_1\dots\alpha_{n-1}}V^{[n]}_{s_n,\alpha_{n-1}}V^{[n-1]}_{\alpha_{n-1}s_{n-1},\alpha_{n-2}}\dots V^{[2]}_{\alpha_2s_2,\alpha_1}V^{[1]}_{\alpha_1s_1}.
\]
Each $V$ now has orthonormal columns and is square or has more rows than columns. The only exception is $V^{[1]}$, which is a column vector, but we can redefine it as an isometry that prepares the two-qubit state described by that column $V^{[1]}\mapsto V^{[1]}(\ket{0}\otimes\ket{0})$. Based on each $V^{[k]}$, it is also convenient to construct an operator $M^{[k]}$, obtained by swapping indices internally within rows and columns as shown in Figure \ref{vandw}.
Finally, we can prepare the matrix product state by sequentially synthesising each isometry $M^{[k]}$ and applying it to the $\ket{0}^{\otimes n}$ state, as shown in Figure \ref{circuit}. Each individual operator is applied to at most $1+\log_2\chi$ qubits.

The fact that we prepare the state starting from the most significant qubit is a consequence of choosing to generate an MPS in the right-canonical form. Using the left-canonical form produces a circuit which prepares the state starting from the least significant qubit \cite{huggins2025efficient}. Using the isometries $M^{[k]}$ as opposed to the isometries $V^{[k]}$ has enabled us to match the most significant qubit in the register to the index $s_{1}$, which indexes the most significant qubit of the target state. This ensures that all qubits in the circuit diagram are represented with horizontal lines, with no need for ancilla qubits or vertical rerouting to match indices between operators. Using this specific construction of the MPS to approximate a DPSS for QPE is also beneficial because it provides compatibility with the semi-classical quantum Fourier transform.

\subsection{The semi-classical Fourier transform}
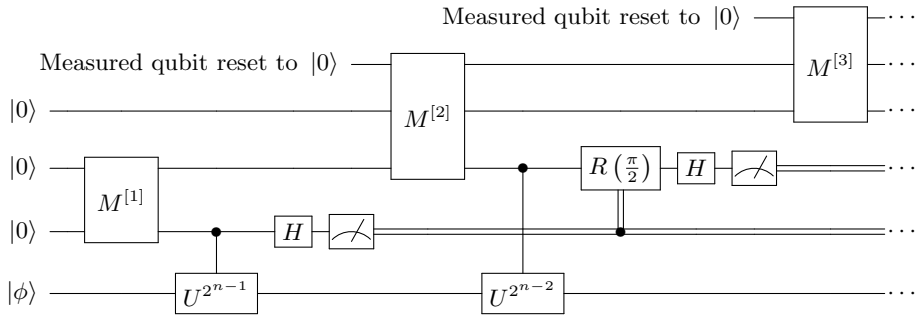
\begin{figure}[tbh]
    \centering
    \[
\Qcircuit @R=1em @C=.3em {
&&&&&&&&&&&&\lstick{\text{Measured qubit reset to }\ket{0}}&\multigate{2}{M^{[3]}}&\qw&\quad\cdots\\
&&&&&&\lstick{\text{Measured qubit reset to }\ket{0}}&\multigate{2}{M^{[2]}}&\qw&\qw&\qw&\qw&\qw&\ghost{M^{[3]}}&\qw&\quad\cdots\\
\lstick{\ket{0}}&\qw&\qw&\qw&\qw&\qw&\qw&\ghost{M^{[2]}}&\qw&\qw&\qw&\qw&\qw&\ghost{M^{[3]}}&\qw&\quad\cdots\\
\lstick{\ket{0}}&\qw&\multigate{1}{M^{[1]}}&\qw&\qw&\qw&\qw&\ghost{M^{[2]}}&\ctrl{2}&\gate{\b[P(4\tilde{\phi})]}&\gate{P\left(\frac{\pi}{2}\right)}&\gate{H}&\meter&\cw&\cw&\quad\cdots\\
\lstick{\ket{0}}&\qw&\ghost{M^{[1]}}&\ctrl{1}&\gate{\b[P(2\tilde{\phi})]}&\gate{H}&\meter&\cw&\cw&\cw&\cctrl{-1}&\cw&\cw&\cw&\cw&\quad\cdots\\
\lstick{\ket{\phi}}&\qw&\qw&\gate{U^{2^{n-1}}}&\qw&\qw&\qw&\qw&\gate{U^{2^{n-2}}}&\qw&\qw&\qw&\qw&\qw&\qw&\quad\cdots
}
    \nonumber
    \]
    \caption{Circuit for performing QPE for a unitary operator $U$ and its eigenstate $\ket{\phi}$ using the semi-classical Fourier transform. \b[An MPS with $\chi=4$ is being used to prepare a control state with dimension $D=2^n$. Since qubits can be measured and recycled, no more than 3 qubits are needed for the control register at any given time. The circuit depicts the measurement of the first two least significant bits of the phase estimate, and it continues until all $n$ bits have been obtained.]}
    \label{circuit2}
\end{figure}

The semi-classical Fourier transform is a variant of the quantum Fourier transform that uses measurements and classically controlled \b[phase shifts] \cite{griffiths1996semiclassical}. This can be beneficial because once a qubit is measured, it can be recycled for other purposes. Using this in QPE necessitates that the most significant qubit of the control state is measured first, yielding the least significant bit of the phase estimate. In Figure \ref{circuit2} we draw the first part of the circuit for performing QPE, incorporating the MPS preparation circuit of Figure \ref{circuit} and the semi-classical Fourier transform implemented with sequential measurements and classically controlled \b[phase shift gates]. \b[The phase shift gates $P(2^k\tilde{\phi})$ that are applied immediately after each of the controlled $U$ operators implement the addition of the random phase discussed in Section \ref{qpe}.]

The circuit in Figure \ref{circuit2} uses an MPS with $\chi=4$, meaning each $M^{[k]}$ acts on at most three qubits. The circuit has structure analogous to the circuits of Ref.~\cite{najafi2023optimum} which use an MPS with $\chi=2$. \b[The dimension of the state that the MPS is approximating is $D=2^n$, but] instead of allocating $n$ qubits to a control register and preparing the entire state on it at once, we can delay the application of each $M^{[k]}$ until a previously used qubit is made available after a measurement. The pattern of measuring out a qubit before recycling it to apply the next operator in the MPS preparation continues until all bits of the phase estimate have been obtained through measurements. The procedure for determining the classically controlled phase is described in detail in Ref.~\cite{griffiths1996semiclassical}, which we will not repeat here.

\subsection{Preparation of MPS with arbitrary dimension}\label{paddingsection}

\begin{figure}[tbh]
    \centering
    \[
\Qcircuit @R=1em @C=.7em {
\lstick{\ket{0}}&\qw&\qw&\targ&\qw&\rstick{\bra{0}}\qw\\
\lstick{\ket{\psi_\text{DPSS}}'}&{/^n}\qw&\gate{-P}&\gate{\geq D}\qwx&\qw&\rstick{\ket{\psi_\text{DPSS}}}\qw
}
    \nonumber
    \]
    \caption{Circuit for preparing a $D$-dimensional DPSS state on $n$ qubits. The $-P$ operator subtracts $P$ from the register and the $\geq D$ operator flips the ancilla qubit if the value in the bottom register is greater than or equal to $D$. The preparation succeeds if the state $\ket{0}$ is measured on the ancilla qubit.}
    \label{ineq}
\end{figure}
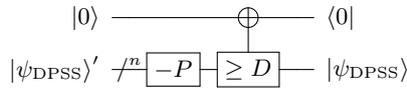

In Section \ref{qpe} we explained how it can be beneficial to prepare a state which has a dimension that is not a power of two. Here we will explain how to prepare an MPS approximation of such a state. First, we follow the procedure described in Section \ref{prep} but use a different target state which is a padded version of the original one. Namely, if we have a $D$-dimensional DPSS state $\ket{\psi_\text{DPSS}}=\sum_{k=0}^{D-1}\gamma_k\ket{k}$ that we are preparing on an $n=\log_2N=\lceil\log_2D\rceil$ qubit register, then we construct
\[
\ket{\psi_\text{DPSS}}'=\sqrt{\frac{1}{\lambda}}\left(\sum_{j=0}^{P-1}\mu_j\ket{j}+\sum_{k=P}^{P+D-1}\gamma_k\ket{k}+\sum_{l=P+D}^{N-1}\nu_l\ket{l}\right),
\]
where $P$ is an integer which determines how the padding is distributed on either side of the state, $\mu_j$ and $\nu_l$ are padding amplitudes that we can choose, and $\lambda$ is a \b[positive] normalisation constant. Once an MPS approximation of this state has been prepared using the circuit in Figure \ref{circuit}, we apply an inequality test to retrieve the DPSS state as shown in Figure \ref{ineq}.

Assuming that the original coefficients were normalised $\sum_k\gamma_k^2=1$, the success probability of the circuit is $|1/\lambda|$. Although producing a state with $\mu_j=\nu_l=0$ would provide a $100\%$ success rate, using such a padding creates a sharp transition in the amplitudes that cannot be approximated well with an MPS. A better objective is to choose the padding amplitudes such that the MPS approximates the DPSS state amplitudes $\gamma_k$ as well as possible while maintaining a reasonably high success rate. In the event of a state preparation failure, flagged by measuring the $\ket{1}$ state on the ancilla qubit, the preparation can be repeated until success.

\begin{figure}[tbh]
    \centering
    \includegraphics[width=\linewidth]{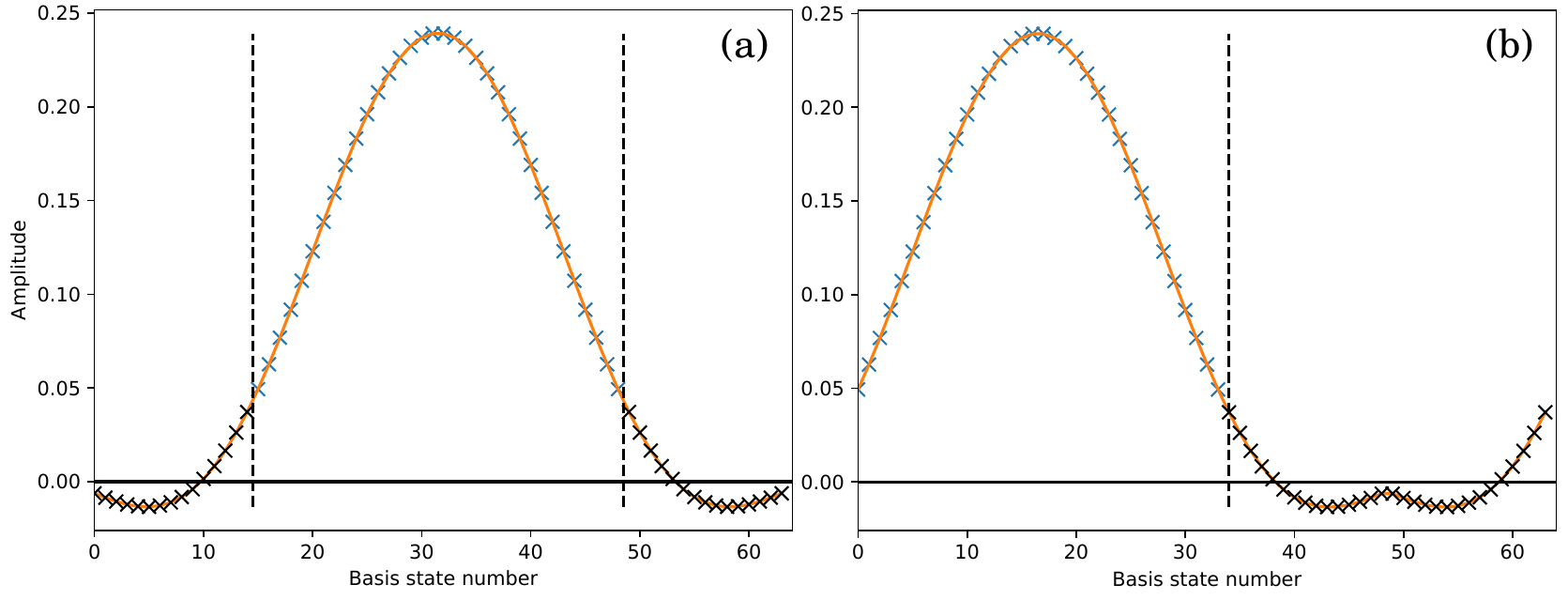}
    \caption{Plot (a) shows the state $\ket{\psi_\text{DPSS}}'$, which is approximated with an MPS. Blue crosses between the black vertical dashed lines indicate the values of $\gamma_k$ for DPSS state with dimension $D=34$, configured for confidence level $1-\delta=99\%$. The orange solid line is the prolate spheroidal wave function, from which the padding amplitudes $\mu_j$ and $\nu_l$ can be sampled, indicated with black crosses. Plot (b) shows the state after the circuit in Figure \ref{ineq} has been applied. In the event of a success, the amplitudes beyond the black dashed line are set to zero by the inequality test, retrieving the desired state $\ket{\psi_\text{DPSS}}$. In this example, the probability for success is $\b[1/\lambda]\approx99.3\%$. The values of the amplitudes in both plots are normalised with $\sum_k\gamma_k^2=1$.}
    \label{fig:padding}
\end{figure}

A suitable choice of padding for even $D$ can be obtained by setting $P=(N-D)/2$ and extending the DPSS in both directions, based on the extended definition of the prolate spheroidal wave function \cite{prolate}, which is the continuous limit of the DPSS as $D\rightarrow\infty$. In the Digital Library of Mathematical Functions \cite{NIST:DLMF}, this is listed with the notation $\text{Ps}_n^m(x,\gamma^2)$, and the function values that we seek correspond to $\text{Ps}_0^0(x/N,dN/2)$, at values of $x$ outside the region $-1\leq x\leq 1$. Figure \ref{fig:padding} shows the values of the padding amplitudes for $D=34$ and $n=6$ as an example.

These values can be numerically obtained by using the \texttt{SpheroidalPS} function available in Mathematica \cite{Mathematica}. At higher values of $D$, determining the padding values with Mathematica becomes the computational bottleneck. This can be circumvented by performing a coarser sampling of padding values and interpolating between the results.
We use this technique in the results shown in Section \ref{results} for states on 17 qubits and beyond. Such an interpolation is an acceptable approximation because ultimately the padding amplitudes are set to zero anyway by the inequality test, and their precise value is unimportant.

Once the state $\ket{\psi_\text{DPSS}}$ has been prepared on the control register, we can proceed to apply the unary iteration shown in Figure \ref{unary} and finally perform an inverse quantum Fourier transform and measurement. Although we can also use the semi-classical Fourier transform here, it is important to note that we cannot recycle qubits as shown in Figure \ref{circuit2} in this case, because both the inequality test and the unary iteration require access to the entire state. 

Using a DPSS state with a dimension that is a power of two enables phase estimation with few qubits, whereas selecting a dimension that is not a power of two can reduce circuit complexity by up to half. Therefore, using a dimension that is a power of two is particularly useful for enabling the use of early-generation fault-tolerant quantum computers where the number of available qubits is severely limited. More advanced quantum computers, where qubit availability is not a limitation, will benefit from faster runtime achieved when using a lower dimension, which does not necessarily have to be a power of two.

\section{QPE with an MPS approximation of a DPSS}\label{results}
We now turn to the practical implementation of DPSS states using MPS approximations. Our analysis pursues three primary objectives: First, we determine the minimum bond dimension required to accurately represent DPSS states across a range of system sizes and parameter values. Second, we examine how the approximation quality varies with state dimension and evaluate how the approximation error introduced by the MPS representation propagates into the phase estimation protocol itself. These investigations establish the conditions under which MPS approximations of DPSS states can serve as effective surrogates for their exact counterparts in QPE. Third, we explain in detail how to generate the MPS that achieves the optimal confidence interval for a given confidence level and control register size, and showcase the result with a representative example.

\subsection{Selecting bond dimension}
The bond dimension of an MPS determines both its representational capacity and its implementation cost, making the selection of an appropriate bond dimension crucial for practical applications. 
We begin by evaluating the fidelity between MPS approximations and target DPSS states across different bond dimensions. Fidelity provides a direct mathematical measure of approximation quality. 
The definition of fidelity $\mathcal{F}$ between these two states is
\[
\mathcal{F}=|\braket{\psi_\text{DPSS}|\psi_\text{MPS}}|^2\label{fidelity}.
\]

For calculating the amplitudes of the DPSS, we use the publicly available implementation in SciPy \cite{2020SciPy-NMeth}, which is based on the tridiagonal eigenvector formulation introduced in Ref.~\cite{slepian1978prolate}. The MPS is constructed based on the target state as described in Section \ref{prep}. This requires fewer classical computational resources than calculating the DPSS state. All of these tasks can be completed in less than a minute on a desktop computer for up to 24 qubits. \b[More details on classical compute time are provided in Appendix \ref{timeappendix}.]

To determine the appropriate bond dimension for approximating DPSS states, we calculated the fidelity using Eq.~\eqref{fidelity} for MPS approximations with bond dimension $2, 4, 8$ and $16$. The results for bond dimensions $2$ and $4$ are tabulated in Table \ref{tab1}. The fidelity increases sharply with bond dimension; for bond dimension $8$ and above, the fidelity is equal to $1$ up to numerical precision for all parameter values evaluated in the table. For 5 qubits or less, the MPS with bond dimension 4 also has perfect fidelity, which is due to the fact that an exact representation of any 5 qubit state is possible with bond dimension 4. Beyond that, when restricted to powers of two, the dimension of the state has almost no impact on the fidelity of the approximation.

\begin{table}[tbh]
    \centering
    \resizebox{\linewidth}{!}{%
    \begin{tabular}{|c|c|c|c|c|c|c|}
    \hline
         \multirow{2}{*}{$n$}&\multicolumn{3}{c|}{Bond dimension 2}&\multicolumn{3}{c|}{Bond dimension 4}  \\
        \cline{2-7}
        &$\delta_\text{DPSS}=10^{-2}$&$\delta_\text{DPSS}=10^{-3}$&$\delta_\text{DPSS}=10^{-4}$&$\delta_\text{DPSS}=10^{-2}$&$\delta_\text{DPSS}=10^{-3}$&$\delta_\text{DPSS}=10^{-4}$  \\
\hline
5&\b[$1.27\times 10^{-4}$] & \b[$5.61\times 10^{-4}$] & \b[$1.16\times 10^{-3}$] & 0 & 0 & 0 \\
\hline
6&\b[$1.35\times 10^{-4}$] & \b[$5.95\times 10^{-4}$] & \b[$1.24\times 10^{-3}$] & \b[$3.38\times 10^{-14}$] & \b[$2.53\times 10^{-12}$] & \b[$4.22\times 10^{-11}$] \\
\hline
7&\b[$1.37\times 10^{-4}$] & \b[$6.04\times 10^{-4}$] & \b[$1.25\times 10^{-3}$] & \b[$4.93\times 10^{-14}$] & \b[$3.78\times 10^{-12}$] & \b[$6.28\times 10^{-11}$] \\
\hline
8&\b[$1.37\times 10^{-4}$] & \b[$6.06\times 10^{-4}$] & \b[$1.26\times 10^{-3}$] & \b[$5.44\times 10^{-14}$] & \b[$4.13\times 10^{-12}$] & \b[$6.88\times 10^{-11}$] \\
\hline
9&\b[$1.37\times 10^{-4}$] & \b[$6.07\times 10^{-4}$] & \b[$1.26\times 10^{-3}$] & \b[$5.51\times 10^{-14}$] & \b[$4.23\times 10^{-12}$] & \b[$7.03\times 10^{-11}$] \\
\hline
10&\b[$1.37\times 10^{-4}$] & \b[$6.07\times 10^{-4}$] & \b[$1.26\times 10^{-3}$] & \b[$5.51\times 10^{-14}$] & \b[$4.25\times 10^{-12}$] & \b[$7.07\times 10^{-11}$] \\
\hline
11&\b[$1.37\times 10^{-4}$] & \b[$6.07\times 10^{-4}$] & \b[$1.26\times 10^{-3}$] & \b[$5.77\times 10^{-14}$] & \b[$4.25\times 10^{-12}$] & \b[$7.08\times 10^{-11}$] \\
\hline
12&\b[$1.37\times 10^{-4}$] & \b[$6.07\times 10^{-4}$] & \b[$1.26\times 10^{-3}$] & \b[$5.60\times 10^{-14}$] & \b[$4.26\times 10^{-12}$] & \b[$7.08\times 10^{-11}$] \\
\hline
13&\b[$1.37\times 10^{-4}$] & \b[$6.07\times 10^{-4}$] & \b[$1.26\times 10^{-3}$] & \b[$5.57\times 10^{-14}$] & \b[$4.26\times 10^{-12}$] & \b[$7.08\times 10^{-11}$] \\
\hline
14&\b[$1.37\times 10^{-4}$] & \b[$6.07\times 10^{-4}$] & \b[$1.26\times 10^{-3}$] & \b[$5.66\times 10^{-14}$] & \b[$4.26\times 10^{-12}$] & \b[$7.08\times 10^{-11}$] \\
\hline
15&\b[$1.37\times 10^{-4}$] & \b[$6.07\times 10^{-4}$] & \b[$1.26\times 10^{-3}$] & \b[$5.44\times 10^{-14}$] & \b[$4.26\times 10^{-12}$] & \b[$7.08\times 10^{-11}$] \\
\hline
16&\b[$1.37\times 10^{-4}$] & \b[$6.07\times 10^{-4}$] & \b[$1.26\times 10^{-3}$] & \b[$5.82\times 10^{-14}$] & \b[$4.26\times 10^{-12}$] & \b[$7.08\times 10^{-11}$] \\
\hline
17&\b[$1.37\times 10^{-4}$] & \b[$6.07\times 10^{-4}$] & \b[$1.26\times 10^{-3}$] & \b[$5.51\times 10^{-14}$] & \b[$4.25\times 10^{-12}$] & \b[$7.08\times 10^{-11}$] \\
\hline
18&\b[$1.37\times 10^{-4}$] & \b[$6.07\times 10^{-4}$] & \b[$1.26\times 10^{-3}$] & \b[$5.53\times 10^{-14}$] & \b[$4.25\times 10^{-12}$] & \b[$7.08\times 10^{-11}$] \\
\hline
19&\b[$1.37\times 10^{-4}$] & \b[$6.07\times 10^{-4}$] & \b[$1.26\times 10^{-3}$] & \b[$5.55\times 10^{-14}$] & \b[$4.26\times 10^{-12}$] & \b[$7.08\times 10^{-11}$] \\
\hline
20&\b[$1.37\times 10^{-4}$] & \b[$6.07\times 10^{-4}$] & \b[$1.26\times 10^{-3}$] & \b[$5.37\times 10^{-14}$] & \b[$4.26\times 10^{-12}$] & \b[$7.09\times 10^{-11}$] \\
\hline
21&\b[$1.37\times 10^{-4}$] & \b[$6.07\times 10^{-4}$] & \b[$1.26\times 10^{-3}$] & \b[$5.71\times 10^{-14}$] & \b[$4.26\times 10^{-12}$] & \b[$7.08\times 10^{-11}$] \\
\hline
22&\b[$1.37\times 10^{-4}$] & \b[$6.07\times 10^{-4}$] & \b[$1.26\times 10^{-3}$] & \b[$5.55\times 10^{-14}$] & \b[$4.22\times 10^{-12}$] & \b[$7.04\times 10^{-11}$] \\
\hline
23&\b[$1.37\times 10^{-4}$] & \b[$6.07\times 10^{-4}$] & \b[$1.26\times 10^{-3}$] & \b[$3.97\times 10^{-14}$] & \b[$4.15\times 10^{-12}$] & \b[$7.21\times 10^{-11}$] \\
        \hline
24&\b[$1.37\times 10^{-4}$] & \b[$6.07\times 10^{-4}$] & \b[$1.26\times 10^{-3}$] & \b[$3.48\times 10^{-13}$] & \b[$4.30\times 10^{-12}$] & \b[$7.29\times 10^{-11}$] \\
\hline
    \end{tabular}%
    }
    \caption{Values of infidelity $1-\mathcal{F}$ for MPS approximations of DPSS states with dimension $2^n$ configured for confidence level $\delta_\text{DPSS}$.}
    \label{tab1}
\end{table}

These results suggest that increasing the bond dimension beyond 8 has no benefit in terms of approximation accuracy, and that the bond dimension of the MPS can be chosen independently of the dimension of the state. 
Since lower bond dimensions reduce implementation cost, we primarily focus on bond dimension 4 for the remainder of this work, as it represents the best balance between implementation efficiency and approximation quality, with bond dimension 2 exhibiting substantially lower fidelity as shown in Table \ref{tab1}.

\subsection{Evaluating impact on phase estimation}\label{impact}
While high fidelity indicates accurate state preparation, it does not directly reveal how MPS approximation errors affect the achievable confidence levels in phase estimation. To assess the practical impact of these approximations, we examine the degradation of confidence levels and confidence intervals of the phase estimate when using MPS-approximated DPSS states.

In phase estimation, if the normalised state on the control register is $\ps=\sum_{k=0}^{D-1}f(k)\ket{k}$ (with integer values of $k$ enumerating the basis states in the control register), then the probability of the phase $\phi$ being within a confidence interval of size $[-d,d]$ can be calculated as \cite{berry2024rapid}:
\[
1-\delta=\frac{d}{\pi}\sum_{j=0}^{D-1}\sum_{k=0}^{D-1}f(j)f(k)\text{sinc}(d(j-k))=\braket{\psi|S|\psi}=\text{Tr}\left[\rho S(d)\right]\label{prob},
\]
where $\text{sinc}(x)\equiv\sin(x)/x$, $\rho=\ps\bra{\psi}$, and $S(d)$ is defined in Eq.~\eqref{sop}. \b[The value of the sinc function at zero (which occurs when $j=k$) is 1, as given by the continuous limit.] The DPSS state minimises $d$ for any given $\delta$, or conversely minimises $\delta$ for any given $d$, because it is by definition the eigenstate of $S(d)$ corresponding to the largest eigenvalue. Any other control state will perform worse, either by producing an increase in $d$, $\delta$, or both. In Section \ref{fixedconf} we analyse in more detail how these variables and the choice of the control state are related when using MPS approximations of DPSS states to achieve a target confidence level.

In this section, we instead focus on a direct comparison between DPSS states and their MPS approximations at fixed values of $d$. Since each DPSS state is an eigenstate of $S(d)$ for a specific value of $d$, we can evaluate the performance of an MPS approximation by fixing this value and calculating the increase in $\delta$ when the DPSS state is replaced with its MPS approximation.

We calculate the relative increase in $\delta$ induced by the MPS approximation as $(\delta_\text{MPS}-\delta_\text{DPSS})/\delta_\text{DPSS}$, where $\delta_\text{MPS}$ and $\delta_\text{DPSS}$ correspond to the confidence levels achieved by the MPS and DPSS state respectively. We evaluate these quantities from Eq.~\eqref{prob}. The classical computational resources required to directly evaluate the double sum in Eq.~\eqref{prob} scale as $\mathcal{O}(D^2)$. This can be done more efficiently by recognising that $S(d)$ is a Toeplitz matrix and that multiplication with a Toeplitz matrix is equal to discrete linear convolution with its first column $s(d)$ \cite{gray2006toeplitz}:
\begin{equation}
\braket{\psi|S|\psi}=\braket{\psi|s(d)*\psi}.
\end{equation}
Fast convolution algorithms based on the discrete Fourier transform and the circular convolution theorem can be used to evaluate this expression with computational complexity $\mathcal{O}(D\log D)$ \cite{oppenheim1999discrete}. 

\begin{figure}[tbh]
    \centering
    \includegraphics[width=0.9\linewidth]{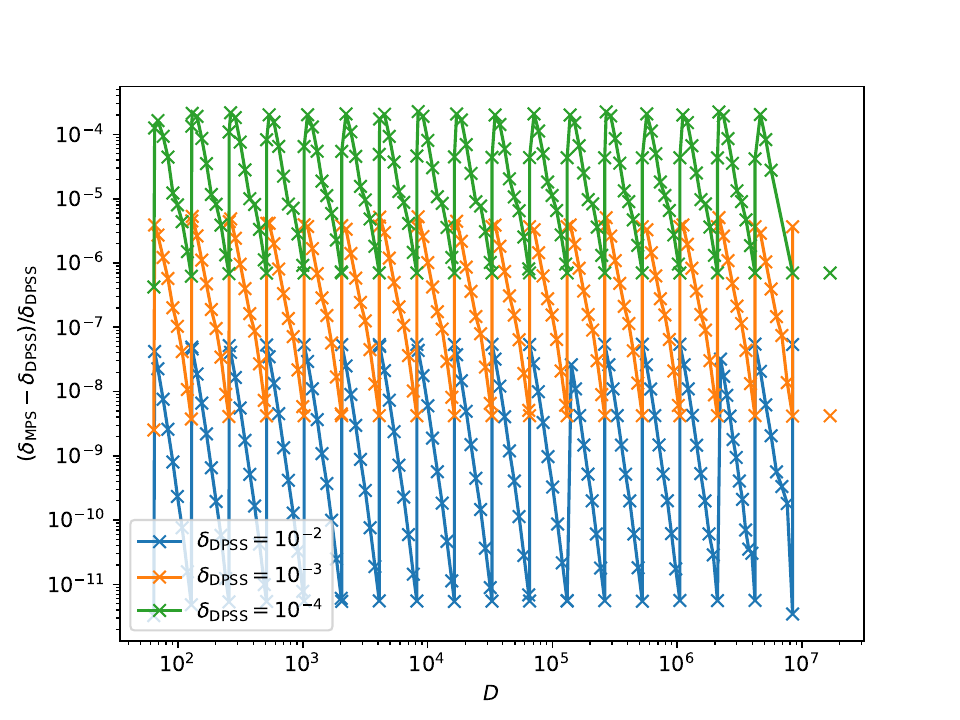}
    \caption{Relative increase in error probability $\delta$ arising from approximating a DPSS state of dimension $D$ with an MPS with bond dimension 4. The minimal values occur at dimensions that are powers of two. Crosses indicate where the dimensions where the error was evaluated, and solid lines interpolate for visual guidance.}
    \label{fig:relerr}
\end{figure}

We plot the relative increase in error probability $(\delta_\text{MPS}-\delta_\text{DPSS})/\delta_\text{DPSS}$ induced by the MPS approximation as a function of the dimension $D$ in Figure \ref{fig:relerr}. For values of $D$ that are not powers of 2, these results use padding values obtained by extending the prolate spheroidal window function beyond the regular domain as explained in Section \ref{paddingsection}. We also tested zero-amplitude padding, but this produces relative errors up to two orders of magnitude larger while exhibiting qualitatively similar behaviour; these results are not shown here.

Overall, the relative increase in error induced by the MPS approximation is well below $0.1\%$ for confidence levels up to $1-10^{-4}$, which is negligible for practical purposes. We attribute this to the fact that the MPS approximates the DPSS state with high fidelity. 

We also observe a repeating pattern in the performance of the MPS approximation, where the maximum value of relative increase in error probability occurs at all values of $D$ that are one larger than a power of two. This gradually improves until the next power of two, after which the error resets to approximately the same level as the previous maximum value.
This indicates that the relative amount of padding matters, with MPS approximation performance improving when padding amplitudes constitute a smaller fraction of the total state. This leaves room for improvement in future work, where it may be possible to find even better padding values to reduce this effect.

\subsection{Convergence with dimension}
The periodic behaviour observed in Figure \ref{fig:relerr} indicates that MPS approximation performance remains approximately constant as the target state dimension doubles, even with fixed bond dimension. We hypothesise that this pattern reflects an inherent property of DPSS states that continues to hold beyond the dimensions we have numerically tested. To support this hypothesis, we provide a theoretical explanation for the periodic behaviour and test whether the underlying mechanism persists as dimension increases.

As $D$ is increased, each DPSS converges to a sequence obtained by sampling the prolate spheroidal wave function at $D$ evenly spaced points. From this perspective, doubling the dimension corresponds to sampling the same underlying function at twice the resolution, inserting new sampling points between each existing pair. This suggests that doubling $D$ requires only refining the least significant qubits while preserving the coarse structure encoded in the more significant qubits. If this property holds consistently, then the periodic pattern should persist for dimensions beyond those in Figure \ref{fig:relerr}.

The structure of the MPS preparation circuit in Figure \ref{circuit} accommodates this refinement naturally: doubling the state dimension introduces new operators exclusively on the least significant qubits. 
This does not necessarily mean that it is optimal to leave all other operators unchanged when new ones are added, but no changes in the previous operators are strictly required to produce refinements only in the least significant qubits. 
Numerically, we find that as qubits are added, previous operators may change noticeably with the new sequence of SVD-s, but this is only indicative of local unitary operators being transferred between neighbouring MPS operators.
To account for this freedom, we will proceed by comparing states directly, rather than comparing the individual operators that are used to prepare them.

To test whether the difference between dimension $D$ and $2D$ states occurs primarily in the least significant qubits, we compare an MPS approximation $\ket{\psi_{(2D)}}=M^{[n+1]}_{(2D)}\dots M^{[1]}_{(2D)}\ket{0}$ of a DPSS with dimension $2D$ to an MPS approximation $\ket{\psi_{(D)}}=M^{[n]}_{(D)}\dots M^{[1]}_{(D)}\ket{0}$ of a DPSS with dimension $D$. We calculate the fidelity $F(D)$ between $\ket{\psi_{(D)}}$ and the reduced state on the $n$ most significant qubits obtained by tracing out the least significant qubit from $\ket{\psi_{(2D)}}$
\[
F(D)=|(\bra{\psi_{(D)}}\otimes\bra{0})\ket{\psi_{(2D)}}|^2+|(\bra{\psi_{(D)}}\otimes\bra{1})\ket{\psi_{(2D)}}|^2.
\]
This expression is obtained by taking the reduced density matrix as
\begin{equation}
    \rho = \braket{0|\psi_{(2D)}}\braket{\psi_{(2D)}|0} + \braket{1|\psi_{(2D)}}\braket{\psi_{(2D)}|1} \, ,
\end{equation}
and then using the expression for the fidelity between a pure state and a mixed state as $\bra{\psi_{(D)}}\rho\ket{\psi_{(D)}}$.

\begin{figure}[tbh]
    \centering
    \includegraphics[width=0.9\linewidth]{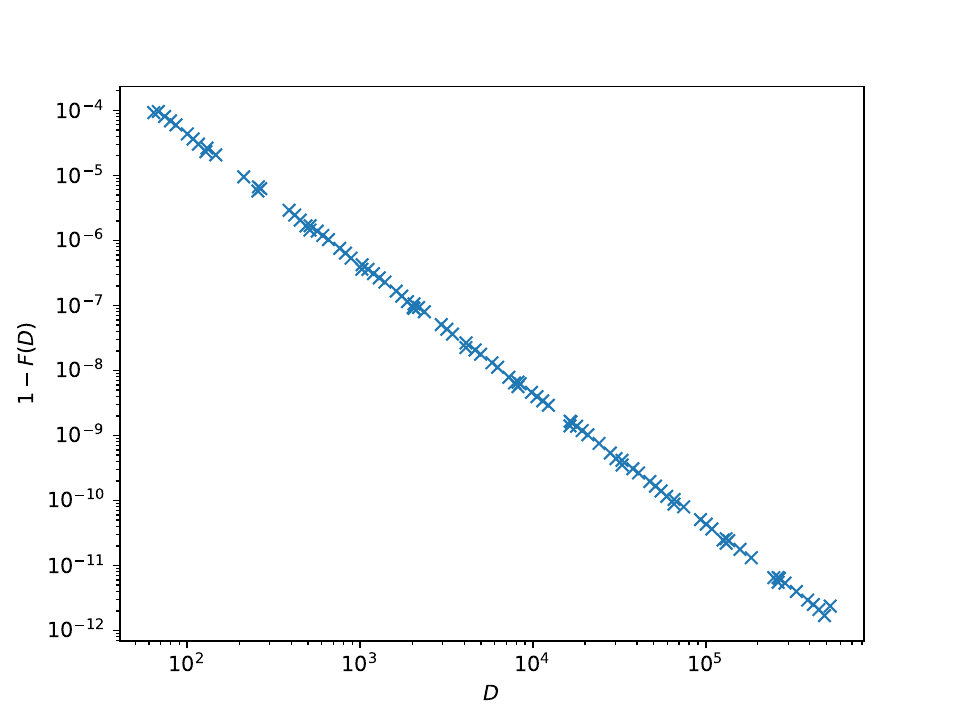}
    \caption{Infidelity between $\ket{\psi_{(D)}}$ and the mixed state left on the $n$ most significant qubits when tracing out the least significant qubit from $\ket{\psi_{(2D)}}$.}
    \label{fig:convergence}
\end{figure}

We plot $1-F(D)$ in Figure \ref{fig:convergence}, and observe that the fidelity is high for any $D$, and increases with $D$. This convergence indicates that the state on the most significant qubits becomes increasingly stable as dimension grows, which is precisely what is required for the periodic pattern to persist at higher dimensions. The observed convergence supports our hypothesis that the pattern will continue beyond the tested range, because the underlying mechanism of refining only the least significant qubits becomes more pronounced rather than degraded as D increases. This provides evidence that the periodic pattern in Figure \ref{fig:relerr} will persist for MPS approximations of DPSS states beyond the dimensions that we have numerically evaluated.

\subsection{Performance at fixed confidence level}\label{fixedconf}
From Figure \ref{fig:relerr}, we have seen that approximating a DPSS state with an MPS produces a small relative decrease in confidence level, but in practical applications, it is desirable to use a procedure which guarantees a fixed confidence level.
As explained in the beginning of Section \ref{impact}, a desired confidence level $1-\delta$ can be achieved either by changing the control state, increasing $d$, or a combination of the two.

\begin{figure}[tbh]
    \centering
    \includegraphics[width=0.9\linewidth]{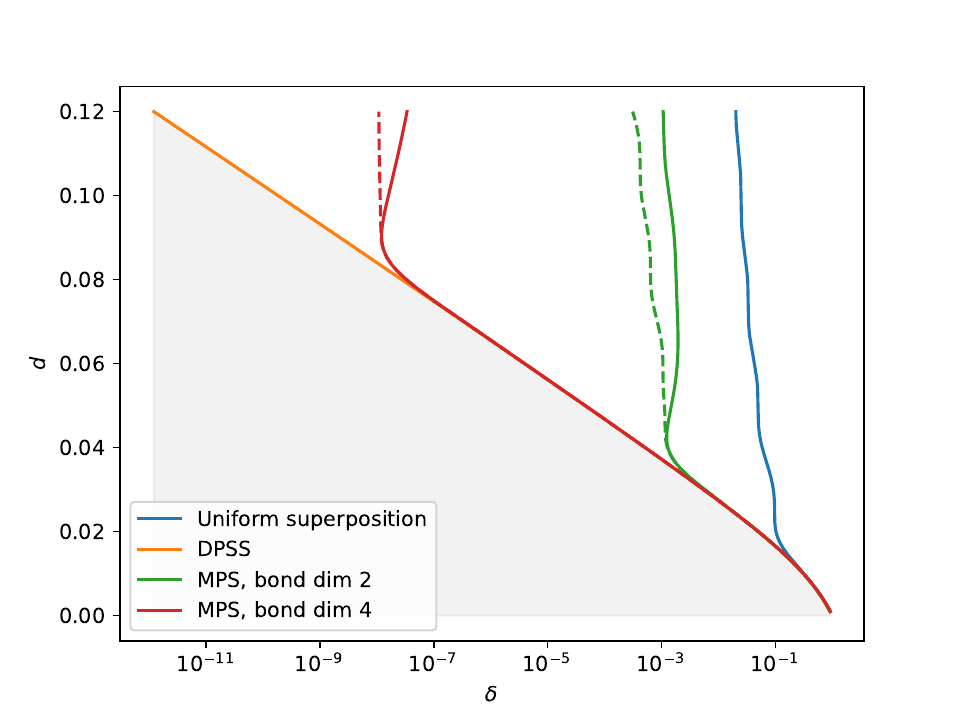}
    \caption{Confidence interval half-width $d$ as a function of error probability $\delta$ for phase estimation using 8 control qubits. The shaded region below the DPSS line is unachievable by any 8 qubit control state and can only be accessed by increasing the dimension of the control state. Solid MPS curves correspond to MPS approximations of DPSS states chosen to be optimal for that value of $d$. Dashed MPS curves indicate the best performing MPS approximations, when optimising the target state over DPSS states configured for alternative values of $d$.}
    \label{interval}
\end{figure}

We will now investigate how the required value of $d$ changes when using different control states to achieve the target confidence level. The control states that we consider are the uniform superposition state, the DPSS state, and MPS approximations of DPSS states. The relation between $d$ and $\delta$ is plotted in Figure \ref{interval} for these types of control states, when the size of the control register is 8 qubits. This serves as a representative example with similar behaviour observed for other control register sizes.

For the uniform superposition, there are no parameters to change for the control state. This means that $\delta$ can only be decreased at the expense of increasing $d$, as shown with the blue curve in Figure \ref{interval}.

For DPSS states, there is a clear optimal choice of how to modify the state and $d$ simultaneously. For any given value of $d$, the optimal choice among DPSS states is the eigenstate of the operator $S(d)$ that corresponds to the largest eigenvalue. The orange curve in Figure \ref{interval} shows the lowest possible value of $d$ for any given $\delta$, which is achieved by using the appropriate DPSS state.
We have shaded the region below this curve to indicate that no other control state of the same dimension can provide the same confidence level at a lower value of $d$. The only way to access this region is by increasing the size of the control register.

For MPSs of bond dimensions 2 and 4, we plot two curves, depending on how the MPS is chosen. For the solid curves, the MPS used on each point in the curve is the approximation of the DPSS state that would be optimal for that value of $d$. At smaller values of $d$, these curves are indistinguishable in the plot, but at higher values of $d$, the MPSs provide confidence levels which are significantly worse than those provided by the corresponding DPSS states. The dashed curves show the best performance achievable by using MPS approximations, when optimizing over different possible target DPSS states, configured for alternative values of $d$.

The results demonstrate that MPS approximations achieve the near-optimal confidence interval width of the ideal DPSS up to a sharp confidence level cutoff that depends on bond dimension. Beyond this cutoff, the MPS cannot maintain optimal performance. Modifying the MPSs by optimising over different target DPSS states slightly improves performance (dashed curves), but still produces a result far from optimal. This is especially pronounced for the $\chi=4$ case, where the dashed curve rises almost vertically beyond $\delta=10^{-8}$, indicating that confidence levels beyond this threshold can only be achieved by substantially increasing $d$. When $d$ approaches $\pi$, the confidence level is guaranteed to approach $100\%$ for any control state.

To understand why the MPS approximations exhibit a sharp performance cutoff at certain confidence levels, we examine how the cutoff relates to approximation fidelity. In Appendix \ref{appa}, we derive an analytic upper bound on the relative increase in $\delta$, as a function of fidelity:
\[
\frac{\delta_\text{MPS}-\delta_\text{DPSS}}{\delta_\text{DPSS}}\leq \left(\sqrt{\mathcal{F}}+\sqrt{\frac{1-\mathcal{F}}{\delta_\text{DPSS}}+\mathcal{F}-1}\right)^2-1.\label{bound}
\]
As the desired confidence level increases ($\delta_\text{DPSS}$ decreases), the bound diverges when $\delta_\text{DPSS}$ becomes comparable to $1-\mathcal{F}$ due to the term $(1-\mathcal{F})/\delta_\text{DPSS}$. This divergence predicts the confidence level regime where the MPS approximation can no longer be guaranteed to perform well, explaining why the cutoff occurs when the desired confidence level approaches the same order of magnitude as the MPS infidelity (see Table \ref{tab1}). Conversely, in the regime $1-\mathcal{F}\ll\delta_\text{DPSS}$, the bound remains small, indicating that high-fidelity MPS approximations produce negligible increases in error probability that require only negligible increases in $d$ for compensation.

Across all bond dimensions tested, MPS-approximated DPSS states substantially outperform the uniform distribution commonly used as the default initial state in quantum phase estimation.

\section{Implementation costs}\label{implementation}
The performance of the MPS preparation circuit is contingent on our ability to synthesise the operators $M^{[k]}$ accurately. One straightforward way of doing this is to extend the isometries to unitary operators and then use one of the many existing methods to perform unitary synthesis \cite{dawson2005solovay, hao2025reducing, amy2013meet, gheorghiu2022t, paradis2024synthetiq, weiden2024high}. A different and possibly more cost-efficient approach would be to use a technique for isometry synthesis \cite{berry2024rapid, lowTradingGatesDirty2024a, itenQuantumCircuitsIsometries2016, szaszNumericalCircuitSynthesis2023}. Reference~\cite{berry2024rapid} describes a custom isometry synthesis scheme designed specifically for MPS preparation, where certain parts between neighbouring operators can be combined. Here, we provide an adapted version of this, reinterpreted as a series of simplifications that arise after performing the cosine-sine decomposition (CSD) \cite{golub2013matrix, paige1994history, sutton2009computing}
on each of the operators. 

\begin{figure}[tbh]
    \centering
    \scalebox{0.95}{$
\begin{aligned}
\Qcircuit @R=1em @C=.7em {
\lstick{\ket{0}}&\qw&\multigate{2}{M^{[k]}}&\qw&\rstick{(\text{a})}\\
\lstick{\ket{0}}&\multigate{2}{M^{[k-1]}}&\ghost{M^{[k]}}&\qw\\
&\ghost{M^{[k-1]}}&\ghost{M^{[k]}}&\qw\\
&\ghost{M^{[k-1]}}&\qw&\qw
}&\qquad\qquad\qquad\qquad\quad\ 
\Qcircuit @R=1em @C=.7em {
\lstick{\ket{0}}&\qw&\qw&\gate{R_y}&\gate{}&\qw&\rstick{(\text{b})}\\
\lstick{\ket{0}}&\multigate{2}{M^{[k-1]}}&\multigate{1}{M^{[k]}_a}&\gate{}\qwx&\multigate{1}{M^{[k]}_b}\qwx&\qw\\
&\ghost{M^{[k-1]}}&\ghost{M^{[k]}_a}&\gate{}\qwx&\ghost{M^{[k]}_b}&\qw\\
&\ghost{M^{[k-1]}}&\qw&\qw&\qw&\qw
}
    \nonumber
    \\~\nonumber\\
\Qcircuit @R=1em @C=.7em {
\lstick{\ket{0}}&\qw&\gate{R_y}&\gate{}&\qw&\rstick{(\text{c})}\\
\lstick{\ket{0}}&\multigate{2}{\widetilde{M}^{[k-1]}}&\gate{}\qwx&\multigate{1}{M^{[k]}_b}\qwx&\qw\\
&\ghost{\widetilde{M}^{[k-1]}}&\gate{}\qwx&\ghost{M^{[k]}_b}&\qw\\
&\ghost{\widetilde{M}^{[k-1]}}&\qw&\qw&\qw
}&\qquad\qquad\qquad
\Qcircuit @R=1em @C=.7em {
\lstick{\ket{0}}&\qw&\gate{R_y}&\qw&\multigate{1}{R_{zy}}&\qw&\qw&\rstick{(\text{d})}\\
\lstick{\ket{0}}&\multigate{2}{\widetilde{M}^{[k-1]}}&\gate{}\qwx&\multigate{1}{M^{[k]}_c}&\ghost{R_{zy}}&\multigate{1}{M^{[k]}_d}&\qw\\
&\ghost{\widetilde{M}^{[k-1]}}&\gate{}\qwx&\ghost{M^{[k]}_c}&\gate{}\qwx&\ghost{M^{[k]}_d}&\qw\\
&\ghost{\widetilde{M}^{[k-1]}}&\qw&\qw&\qw&\qw&\qw
}
\nonumber
\end{aligned}
$}
    \caption{Circuits obtained by applying a series of useful decompositions to facilitate synthesis of an operator $M^{[k]}$ within an MPS preparation circuit. Circuit (a) shows a part of the MPS circuit from Figure \ref{circuit}. Circuit (b) is obtained by applying the CSD to (a). Circuit (c) is obtained by merging $M^{[k-1]}$ with $M{_a^{[k]}}$ in circuit (b). Circuit (d) is obtained by performing a type D Cartan decomposition to $M_b^{[k]}$ in circuit (c). This sequence of decompositions (a)-(d) can be sequentially applied for each value of $k$ to obtain a decomposed circuit for the entire MPS.}
    \label{qsd}
\end{figure}
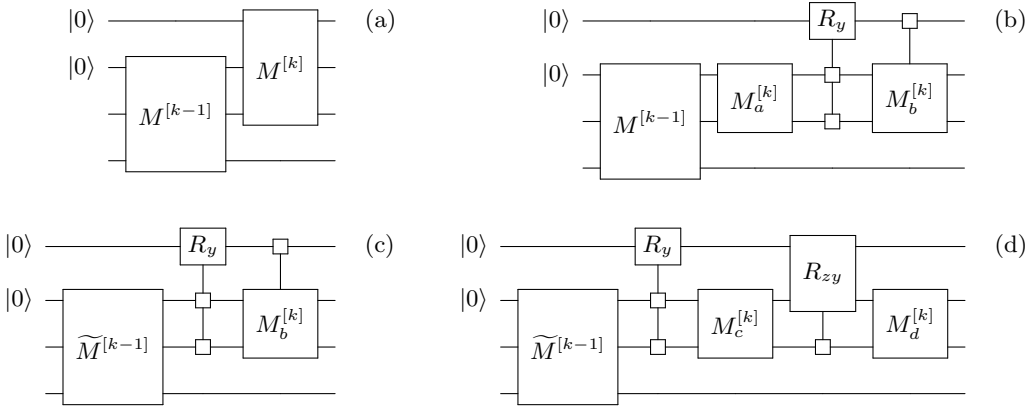

When the bond dimension is 4, all operators $M^{[k]}$ besides $M^{[1]}, M^{[2]}, M^{[n-1]}$ and ${M^{[n]}}$ are isometries which have 8 rows and 4 columns. Splitting the rows in half to define $4\times 4$ sub-matrices $M_0^{[k]}$ and $M_1^{[k]}$, we can use the 2-by-1 CSD, which is a decomposition of the form
\[
\label{2by1csd}
M^{[k]}=
\left(
\begin{array}{c}
M^{[k]}_0\\
M^{[k]}_1
\end{array}
\right)=\left(
\begin{array}{cc}
M^{[k]}_{b,0}&\\
&M^{[k]}_{b,1}
\end{array}
\right)
\left(
\begin{array}{c}
C\\
-S
\end{array}
\right)M^{[k]}_a,
\]
where $M^{[k]}_{b,0}$, $M^{[k]}_{b,1}$, and ${M_a^{[k]}}$ are two-qubit unitaries, and $C$ and $S$ are diagonal matrices with $C^2+S^2=\openone$, which can be implemented in a quantum circuit with a controlled $R_y$ gate. Figure \ref{qsd}(a) shows a part of the MPS circuit from Figure \ref{circuit}, and Figure \ref{qsd}(b) shows the corresponding circuit part when the CSD is applied to $M^{[k]}$. Note that Eq.~\eqref{2by1csd} is presented with the common convention that the $R_y$ gate is applied to the most significant qubit, allowing convenient use of the block-matrix notation. However, this is not a limitation of the 2-by-1 CSD, meaning that the decomposition can be applied to any qubit ordering by switching the appropriate rows and columns before and after applying the decomposition in the form of Eq.~\eqref{2by1csd}. 
In our MPS preparation circuit, we apply the decomposition with the rotation on the least significant qubit to remain compatible with the ordering established in Section \ref{prep}.

We can reduce the cost of the circuit by combining the left two gates to obtain $\widetilde{M}^{[k-1]}=(\openone_2\otimes M_a^{[k]})M^{[k-1]}$ as shown in Figure \ref{qsd}(c). This circuit is equivalent to the construction in Ref.~\cite{berry2024rapid} with the adaptation that it is configured on an MPS circuit that uses no ancilla qubits. To implement the full MPS circuit, we can start by combining $M^{[n]}$ and $M^{[n-1]}$ with $M^{[n-2]}$ to retrieve $\widetilde{M}^{[n-2]}=(\openone_4\otimes M^{[n]})(\openone_2\otimes M^{[n-1]})M^{[n-2]}$. Then, we apply the CSD and merge steps iteratively to obtain $\widetilde{M}^{[n-3]}\dots \widetilde{M}^{[2]}$. We can also apply the CSD to $\widetilde{M^{[2]}}$, but $M_a^{[2]}$ cannot be merged with $M^{[1]}$ because it is a two-qubit operator that is not applied to the same qubits.

Since we are preparing a state with real-valued amplitudes, all of the gates in the circuit implement orthogonal operators. 
Controlled orthogonal operators can be decomposed with a type D Cartan decomposition \cite{wierichs2025recursive}. 
We will apply this to the $M_b^{[k]}$ gates to simplify the circuit further. 
The type D Cartan decomposition can be written as
\[
\left(
\begin{array}{cc}
M^{[k]}_{b,0}&\\
&M^{[k]}_{b,1}
\end{array}
\right)=M_d^{[k]}\left(\begin{array}{cc}
C&S\\
-S&C
\end{array}\right)M_c^{[k]},
\]
where $M_c^{[k]}$ and $M_d^{[k]}$ are two-qubit unitaries and $C^2+S^2=\openone_2$, which can be implemented in a quantum circuit with a uniformly controlled $R_{zy}$ gate. Figure \ref{qsd}(d) shows the result after applying this decomposition to the circuit in Figure \ref{qsd}(c).

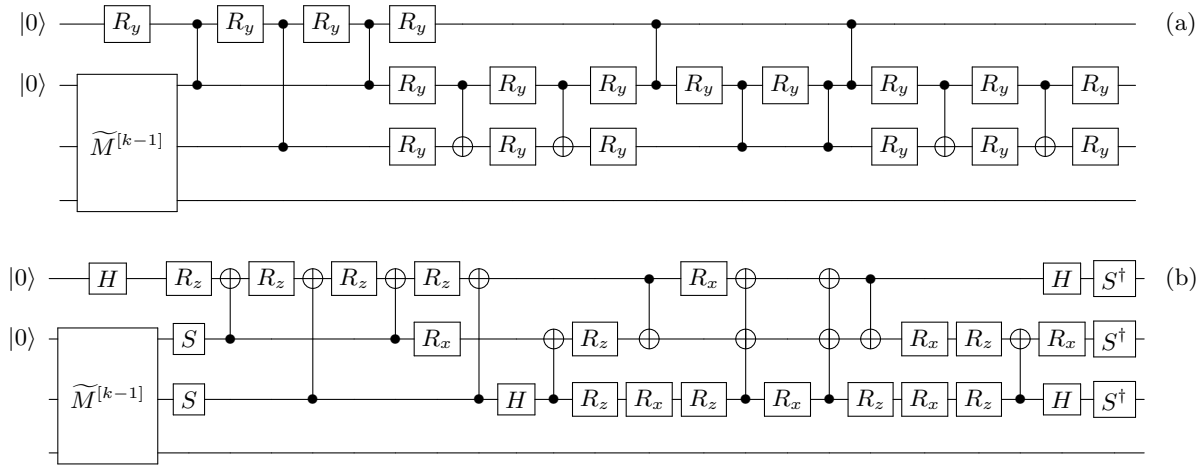
\begin{figure}[tbh]
    \centering
\scalebox{0.8}{$
\begin{aligned}
\Qcircuit @R=1em @C=.7em {
\lstick{\ket{0}}&\gate{R_y}&\ctrl{1}&\gate{R_y}&\ctrl{2}&\gate{R_y}&\ctrl{1}&\gate{R_y}&\qw&\qw&\qw&\qw&\ctrl{1}&\qw&\qw&\qw&\qw&\ctrl{1}&\qw&\qw&\qw&\qw&\qw&\qw&\rstick{(\text{a})}\\
\lstick{\ket{0}}&\multigate{2}{\widetilde{M}^{[k-1]}}&\control\qw&\qw&\qw&\qw&\control\qw&\gate{R_y}&\ctrl{1}&\gate{R_y}&\ctrl{1}&\gate{R_y}&\control\qw&\gate{R_y}&\ctrl{1}&\gate{R_y}&\ctrl{1}&\control\qw&\gate{R_y}&\ctrl{1}&\gate{R_y}&\ctrl{1}&\gate{R_y}&\qw&\\
&\ghost{\widetilde{M}^{[k-1]}}&\qw&\qw&\control\qw&\qw&\qw&\gate{R_y}&\targ&\gate{R_y}&\targ&\gate{R_y}&\qw&\qw&\control\qw&\qw&\control\qw&\qw&\gate{R_y}&\targ&\gate{R_y}&\targ&\gate{R_y}&\qw&\\
&\ghost{\widetilde{M}^{[k-1]}}&\qw&\qw&\qw&\qw&\qw&\qw&\qw&\qw&\qw&\qw&\qw&\qw&\qw&\qw&\qw&\qw&\qw&\qw&\qw&\qw&\qw&\qw&
}\nonumber
\\~\nonumber\\
\Qcircuit @R=1em @C=.35em {
\lstick{\ket{0}}&\gate{H}&\gate{R_z}&\targ&\gate{R_z}&\targ&\gate{R_z}&\targ&\gate{R_z}&\targ&\qw&\qw&\qw&\ctrl{1}&\gate{R_x}&\targ&\qw&\targ&\ctrl{1}&\qw&\qw&\qw&\gate{H}&\gate{S^\dagger}&\qw&\rstick{(\text{b})}\\
\lstick{\ket{0}}&\multigate{2}{\widetilde{M}^{[k-1]}}&\gate{S}&\ctrl{-1}&\qw&\qw&\qw&\ctrl{-1}&\gate{R_x}&\qw&\qw&\targ&\gate{R_z}&\targ&\qw&\targ&\qw&\targ&\targ&\gate{R_x}&\gate{R_z}&\targ&\gate{R_x}&\gate{S^\dagger}&\qw&\\
&\ghost{\widetilde{M}^{[k-1]}}&\gate{S}&\qw&\qw&\ctrl{-2}&\qw&\qw&\qw&\ctrl{-2}&\gate{H}&\ctrl{-1}&\gate{R_z}&\gate{R_x}&\gate{R_z}&\ctrl{-2}&\gate{R_x}&\ctrl{-2}&\gate{R_z}&\gate{R_x}&\gate{R_z}&\ctrl{-1}&\gate{H}&\gate{S^\dagger}&\qw&\\
&\ghost{\widetilde{M}^{[k-1]}}&\qw&\qw&\qw&\qw&\qw&\qw&\qw&\qw&\qw&\qw&\qw&\qw&\qw&\qw&\qw&\qw&\qw&\qw&\qw&\qw&\qw&\qw&\qw&
}\nonumber
\end{aligned}
$}
    \caption{Two possible circuits for synthesising one operator $\widetilde{M}^{[k]}$. Both circuits contain only single qubit rotations and Clifford gates. Circuit (a) is obtained from using known efficient gate implementations of the gates in Figure \ref{qsd}(d). Circuit (b) is obtained by optimising Circuit (a) and contains one fewer rotation gate.}
    \label{qsdcirc}
\end{figure}

Each of the gates in Figure \ref{qsd}(d) has a known efficient implementation. The gates $M_c^{[k]}$ and $M_d^{[k]}$ can be constructed with the optimal decomposition of an orthogonal two-qubit gate \cite{wei2012decomposition}. The controlled rotations can be decomposed with the method of Ref.~\cite{shendeSynthesisQuantumLogic2006}. The result from applying these techniques to each of the gates is shown in Figure \ref{qsdcirc}(a). The rightmost CZ gate of the decomposition of the controlled $R_y$ gate has been omitted, since it can be absorbed into the controlled $M_b^{[k]}$ gate before performing the type D Cartan decomposition.

An alternative circuit with one fewer rotation gate is provided in Figure \ref{qsdcirc}(b), which is obtained by processing the circuit in Figure \ref{qsdcirc}(a) with a ZX-calculus-based circuit optimisation algorithm \cite{kissinger2019pyzx, duncan2020graph}. The algorithm is implemented with publicly available code \cite{pyzx}, provided by the authors of Ref.~\cite{kissinger2019pyzx}. The first $S$ gates in the circuit for $M^{[k]}$ could be removed by combining them with the $S^\dagger$ gates produced in the circuit for $\widetilde{M}^{[k-1]}$, but they are displayed in Figure \ref{qsdcirc}(b) for clarity to convey that the operator $\widetilde{M}^{[k-1]}$ was not modified during the circuit optimisation procedure. 

A numerical solver can be used to determine the angles for each rotation. We tested this approach using the \texttt{minimize} function in Scipy \cite{2020SciPy-NMeth} with default parameters to minimise the Frobenius distance between the desired operator and the operator implemented by the circuits in Figure \ref{qsdcirc}. We find that for both the MPS matrices for a DPSS state, as well as for randomly generated orthogonal matrices, the optimisation procedure is successful, achieving a Frobenius distance below $10^{-10}$ regardless of how the initial values of the angles are chosen. Alternatively, the angles could be directly calculated by explicitly performing the decompositions and subsequent ZX-calculus based optimisation procedure on $\widetilde{M}^{[k]}$. \b[With fixed bond dimension, the classical computational resources needed for calculating the angles scale linearly with $n$ (logarithmically with state dimension), because there are $n$ operators for which the angles need to be determined. This is a negligible cost compared to the SVD and DPSS amplitude calculation routines, which scale exponentially with $n$.]

As a result, the implementation of each $\widetilde{M}^{[k]}$ involves synthesising 17 single-qubit rotations if we use the circuit in Figure \ref{qsdcirc}(b). As special cases, $\widetilde{M}^{[2]}$ can be implemented with 16 rotations (the first two rotations can be combined) and $(\openone_2\otimes M_a^{[2]})(M^{[1]}\otimes \openone_2)$ can be implemented with eight rotations because it performs a three-qubit state preparation \b[with real amplitudes] \cite{perdomo2022preparation, 5bw6-339b}. In total, the state preparation circuit uses $24 + 17(n-4)$ single-qubit rotations. Assuming access to the Clifford+T gate set, single-qubit rotations can be synthesised  probabilistically, with the aid of one ancilla qubit, with an average cost of $\b[0.53] \log_2(1/\epsilon) + 4.86$ T gates \cite{kliuchnikov2023shorter}, where $\epsilon$ is the maximum allowed error in the \b[value of the ]rotation angle. The total number of T gates \b[to prepare an MPS on $n$ qubits] is therefore
\[
\text{T-cost}&\approx \b[9.01]n\log_2(1/\epsilon)+82.62n-\b[23.32]\log_2(1/\epsilon)-213.84\, . \label{tcp}
\]

\b[If the goal is to prepare a DPSS state with dimension $D$, then $n=\lceil\log_2D\rceil$ and if $D$ is not a power of two, then an additional cost of at most $2n$ Toffoli gates is incurred for the subtraction \cite{gidney2018halving} and comparator \cite{vandaele2026asymptotically} operations described in Section \ref{paddingsection}, regardless of the value of $\epsilon$. This extra cost would be present independently of how the MPS is prepared or how the rotations are synthesized].

In Table \ref{tab:cost}, we compare this cost to two different methods of preparing a DPSS state. The first is Grover-Rudolph state preparation \cite{grover2002creating}, which can be used to prepare a state with arbitrary amplitudes. This prepares the exact DPSS state, but is not taking advantage of any properties of the state to reduce cost. We calculate the cost based on the implementation provided in Ref.~\cite{berry2024rapid}, which optimises the usage of advanced QROM \cite{lowTradingGatesDirty2024a} (also known as QROAM \cite{berry2019qubitization}).

The second method that we compare to is the method of preparing the MPS approximation of the DPSS state (with $\chi=4$), using the MPS preparation procedure of Ref.~\cite{berry2024rapid}, which also uses QROM. For both of these competing approaches, the dominant cost is expressed as the number of Toffoli gates used.

\b[The number of T or Toffoli gates is used as a proxy for the overall complexity of the quantum circuits because they are the most difficult gates to implement in the surface code \cite{Fowler_2012}.
The best known technique for implementing these gates in the surface code involves magic state injection, where a high code distance magic state is designed and prepared specifically for the type of gate to be implemented.
For this reason, it is most efficient to avoid implementing Toffolis with T gates, which requires 4 T gates per Toffoli \cite{gidney2018halving}, and instead directly use $\ket{CCZ}$ magic states \cite{kook2026lowspatialcostccz,gidney2019efficient,haah2018codes} for circuits containing Toffoli gates.]

\b[
Currently, the best known method for preparing magic states \cite{gidney2025factor2048bitrsa,low2026denserplanarsurfacecode} entails the following procedure.
First, low code distance $\ket{T}$ magic states are prepared by cultivation \cite{gidney2024magic, claes2025cultivatingtstatessurface, rosenfeld2025magicstatecultivationsuperconducting}.
These are then used to produce one high code distance $\ket{CCZ}$ magic state using $8\ket{T}$ to $\ket{CCZ}$ distillation \cite{Jones_2013}.
This magic state can then be consumed to implement a Toffoli gate.
To implement a T gate, one must further distil two high code distance $\ket{T}$ magic states from the $\ket{CCZ}$ magic state using a catalysed $\ket{CCZ}$ to $2\ket{T}$ conversion \cite{gidney2019efficient,low2026denserplanarsurfacecode}.
Although this means that two T gates can be produced from the same resource state as one Toffoli gate, there is the added cost of conversion and the doubled cost of consumption.
Due to this overhead, one Toffoli gate may be cheaper to implement than two T gates, but the exact ratio will depend on the compilation strategy and the application \cite{low2026denserplanarsurfacecode}.
]

\b[Table \ref{tab:cost} compares the T gate count of our MPS preparation circuit to the Toffoli gate count of] the MPS preparation method of Ref.~\cite{berry2024rapid}. Both MPS implementations provide considerable resource savings when compared to Grover-Rudolph DPSS state preparation at higher dimensions.
Furthermore, both of the methods that we are comparing against use many more ancilla qubits, which are required for storing rotation angle values written by QROM.
It is also important to note that $\epsilon$ describes the error allowed for \b[the values of the angles in] individual rotations. The overall error of the entire circuit is difficult to account for, but generally the total error increases with the number of gates in the circuit. This means that for larger circuits, a more stringent requirement on $\epsilon$ is also required to achieve the same overall error for the entire circuit. From this perspective, a reduced gate count as a function of $n$ has an additional benefit of enabling a more lenient choice of $\epsilon$, which further decreases the number of gates needed in total.
\begin{table}[tbh]
    \centering
    \resizebox{\linewidth}{!}{%
    \begin{tabular}{|c|c|c|c|c|c|c|c|c|}
\hline
\multirow{2}{*}{$n$}&\multirow{2}{*}{Method}&\multirow{2}{*}{Gate type}&\multicolumn{6}{c|}{$\log_2(1/\epsilon)$}\\
        \cline{4-9}
        &&&5&10&15&20&25&30\\
\hline
10&
\begin{tabular}{@{}c@{}}
GR, Refs.~\cite{berry2024rapid, grover2002creating,lowTradingGatesDirty2024a}\\MPS, Ref.~\cite{berry2024rapid}\\MPS, Eq.~\eqref{tcp}\end{tabular} &
\begin{tabular}{@{}c@{}}Toffoli\\Toffoli\\T\end{tabular} &
\begin{tabular}{@{}c@{}}490\\697\\\b[947]\end{tabular} &
\begin{tabular}{@{}c@{}}636\\952\\\b[1281]\end{tabular} &
\begin{tabular}{@{}c@{}}761\\1207\\\b[1615]\end{tabular} &
\begin{tabular}{@{}c@{}}858\\1462\\\b[1948]\end{tabular} &
\begin{tabular}{@{}c@{}}948\\1717\\\b[2282]\end{tabular} &
\begin{tabular}{@{}c@{}}1038\\1972\\\b[2616]\end{tabular} \\
\hline
20&
\begin{tabular}{@{}c@{}}
GR, Refs.~\cite{berry2024rapid, grover2002creating,lowTradingGatesDirty2024a}\\MPS, Ref.~\cite{berry2024rapid}\\MPS, Eq.~\eqref{tcp}\end{tabular} &
\begin{tabular}{@{}c@{}}Toffoli\\Toffoli\\T\end{tabular} &
\begin{tabular}{@{}c@{}}16362\\1497\\\b[2223]\end{tabular} &
\begin{tabular}{@{}c@{}}20972\\2052\\\b[3008]\end{tabular} &
\begin{tabular}{@{}c@{}}24817\\2607\\\b[3792]\end{tabular} &
\begin{tabular}{@{}c@{}}27642\\3162\\\b[4577]\end{tabular} &
\begin{tabular}{@{}c@{}}30212\\3717\\\b[5361]\end{tabular} &
\begin{tabular}{@{}c@{}}32782\\4272\\\b[6145]\end{tabular} \\
\hline
30&
\begin{tabular}{@{}c@{}}
GR, Refs.~\cite{berry2024rapid, grover2002creating,lowTradingGatesDirty2024a}\\MPS, Ref.~\cite{berry2024rapid}\\MPS, Eq.~\eqref{tcp}\end{tabular} &
\begin{tabular}{@{}c@{}}Toffoli\\Toffoli\\T\end{tabular} &
\begin{tabular}{@{}c@{}}524266\\2297\\\b[3500]\end{tabular} &
\begin{tabular}{@{}c@{}}671724\\3152\\\b[4735]\end{tabular} &
\begin{tabular}{@{}c@{}}794609\\4007\\\b[5970]\end{tabular} &
\begin{tabular}{@{}c@{}}884730\\4862\\\b[7205]\end{tabular} &
\begin{tabular}{@{}c@{}}966660\\5717\\\b[8440]\end{tabular} &
\begin{tabular}{@{}c@{}}1048590\\6572\\\b[9675]\end{tabular} \\
\hline
\end{tabular}%
}
    \caption{Cost of state preparation for a range of values of $n$ and $\epsilon$. In each table entry, the top number is the number of Toffoli gates needed to prepare the exact DPSS state using a variant of arbitrary state preparation based on the Grover-Rudolph method \cite{grover2002creating, lowTradingGatesDirty2024a} that optimises the usage of QROM \cite{berry2024rapid}. The middle number is the number of Toffoli gates \b[needed] to prepare an MPS with $\chi=4$ using the method of Ref.~\cite{berry2024rapid}. The bottom number is the number of T-gates needed for preparing an MPS with $\chi=4$ using the implementation given in Figure \ref{qsdcirc}(b), with the cost calculated from Eq.~\eqref{tcp}.}
    \label{tab:cost}
\end{table}

\b[It may also be of practical interest to estimate the number of T gates required to prepare a DPSS state that achieves a desired target confidence interval half-width $d$ at a given confidence level $1-\delta$. To do this, we can use the asymptotic expression that relates the required DPSS state dimension $D$ to $\delta$ and $d$ \cite{berry2024rapid}:
\[
D\simeq\frac{\ln(1/\delta)}{2d}+\frac{\ln(8\pi\ln(4\sqrt{\pi}/\delta))}{4d}+\mathcal{O}\left(\frac{\ln\ln(1/\delta)}{d\ln(1/\delta)}\right).
\]
A more precise value can be obtained by numerically searching for the value of $D$ that produces the desired value of $\delta$ at a given value of $d$ in Eq.~\eqref{prob}. This dimension may need to be further increased slightly, to compensate for the finite value of $\epsilon$. Once the dimension has been determined, the cost can be calculated with Eq.~\eqref{tcp}.]

\b[Table \ref{tab:dpss_vs_textbook} shows the value of $D$ for a variety of values of $\delta$ and $d$, calculated from Eq.~\eqref{prob}. We provide a comparison to the textbook control state, as well as the sine window control state. At higher confidence levels, the difference in performance between the control states becomes increasingly pronounced.]

\begin{table}[tbh]
    \centering
    \resizebox{\linewidth}{!}{%
    \b[\begin{tabular}{|c|c|c|c|c|c|c|c|}
\hline
\multirow{2}{*}{Confidence level}&\multirow{2}{*}{Control state}&\multicolumn{6}{c|}{Confidence interval half-width (radians)}\\
        \cline{3-8}
        &&0.2&0.1&0.05&0.01&0.005&0.001\\
\hline
95\%&
\begin{tabular}{@{}c@{}}DPSS\\Sine\\Uniform\end{tabular} &
\begin{tabular}{@{}c@{}}26\\29\\61\end{tabular} &
\begin{tabular}{@{}c@{}}52\\58\\129\end{tabular} &
\begin{tabular}{@{}c@{}}103\\115\\260\end{tabular} &
\begin{tabular}{@{}c@{}}513\\573\\1303\end{tabular} &
\begin{tabular}{@{}c@{}}1026\\1146\\2605\end{tabular} &
\begin{tabular}{@{}c@{}}5127\\5727\\13025\end{tabular} \\
\hline
99\%&
\begin{tabular}{@{}c@{}}DPSS\\Sine\\Uniform\end{tabular} &
\begin{tabular}{@{}c@{}}36\\38\\323\end{tabular} &
\begin{tabular}{@{}c@{}}71\\75\\646\end{tabular} &
\begin{tabular}{@{}c@{}}141\\149\\1293\end{tabular} &
\begin{tabular}{@{}c@{}}702\\743\\6463\end{tabular} &
\begin{tabular}{@{}c@{}}1403\\1486\\12926\end{tabular} &
\begin{tabular}{@{}c@{}}7013\\7426\\64628\end{tabular} \\
\hline
99.9\%&
\begin{tabular}{@{}c@{}}DPSS\\Sine\\Uniform\end{tabular} &
\begin{tabular}{@{}c@{}}48\\86\\3169\end{tabular} &
\begin{tabular}{@{}c@{}}96\\172\\6369\end{tabular} &
\begin{tabular}{@{}c@{}}191\\344\\12742\end{tabular} &
\begin{tabular}{@{}c@{}}954\\1719\\63717\end{tabular} &
\begin{tabular}{@{}c@{}}1907\\3438\\127434\end{tabular} &
\begin{tabular}{@{}c@{}}9535\\17187\\637167\end{tabular} \\
\hline
99.99\%&
\begin{tabular}{@{}c@{}}DPSS\\Sine\\Uniform\end{tabular} &
\begin{tabular}{@{}c@{}}60\\177\\31721\end{tabular} &
\begin{tabular}{@{}c@{}}120\\356\\63614\end{tabular} &
\begin{tabular}{@{}c@{}}240\\712\\127304\end{tabular} &
\begin{tabular}{@{}c@{}}1198\\3561\\636700\end{tabular} &
\begin{tabular}{@{}c@{}}2396\\7122\\1273404\end{tabular} &
\begin{tabular}{@{}c@{}}11980\\35608\\6367028\end{tabular} \\
\hline
\end{tabular}]%
}
    \caption{\b[Control state dimension $D$ required to achieve a given confidence level and interval width.]}
    \label{tab:dpss_vs_textbook}
\end{table}

\b[Besides the control state preparation costs described in this section, there are additional costs for performing phase estimation. If unary iteration is used as described in Section \ref{qpe}, then $D$ applications to controlled $U$ are used, as well as $D-2$ Toffoli gates for the iteration procedure itself \cite{babbush2018encoding}. If unary iteration is not used, then $\lceil2^{\log_2D}\rceil$ applications of controlled $U$ are required.
One phase shift gate per qubit is required for adding the randomly selected phase discussed in Section \ref{qpe}.
The semi-classical Fourier transform requires at most one phase shift gate per qubit (depending on the measurement outcome), but this incurs no additional cost, because it can be combined with the phase shift gate used for the random phase that directly precedes it in the circuit, as can be seen in Figure \ref{circuit2}.]

\b[
The cost of implementing the controlled $U$ operations will depend entirely on the application, but it is expected to incur the majority of the total cost of phase estimation. In many cases the cost of performing the controlled unitary is similar to the cost of performing the unitary without the control. For example, when using qubitization, the effect of controlling $U$ can be achieved by controlling a reflection instead, incurring an overhead of just one Toffoli gate \cite{babbush2018encoding}.]

The application of the decompositions shown in Figure \ref{qsd} generalises to the case of higher bond dimension and can be used to systematically take advantage of the sequential structure of an MPS. This procedure can be used for preparing the MPS approximation of any state with real-valued amplitudes, making it also applicable to state preparation problems that are not necessarily connected to QPE or DPSS states.

\section{Conclusion}
In this work, we have demonstrated that a DPSS state can be prepared with high fidelity using an MPS approximation with bond dimension 4.
The DPSS provides phase estimates with optimal performance in terms of the confidence interval.
In practical applications, the confidence level is typically no more than 99\%.
In that case, the relative increase in the error is no more than one part in $10^7$.
Even if the required confidence level is 99.99\%, the relative increase in the error is still no more than $3\times10^{-4}$.
\b[Since the relative increase in error probability is negligibly small, we have shown that our MPS approximation provides performance which is indistinguishable from the optimal DPSS result in a practical setting.]

Our methods can be applied to DPSS states of any dimension, but they have additional benefits if the dimension is a power of 2.
First, the accuracy of the approximation is very high.
If the confidence level is 99\%, then the relative increase in error is no more than one part in $10^{11}$.
Second, the state preparation circuit is compatible with the semi-classical Fourier transform.
This enables the state to be generated as it is used for the phase estimation, so that no more than three qubits are needed for the control at once.
This is important in situations where the quantum computer has a very limited number of logical qubits.

Our analysis of the performance of the MPS approximation is numerical, and so limited to a finite set of dimensions, but the features of the results strongly indicate that the MPS approximation is accurate for any dimension.
We observe periodic behaviour with the dimension, where the performance of the MPS approximation is very similar as the dimension is increased by a factor of 2.
The performance is best for dimensions that are a power of 2, and worst if the dimension is one more than a power of 2.
As the dimension is increased, the performance improves until it is best again at the next power of 2.

We expect that this behaviour is because doubling the dimension corresponds to approximating the same function, but with one extra less-significant bit, so only a small adjustment to the state is needed.
To support this explanation, we showed that the state prepared by the MPS circuit converges as the dimension of the DPSS is increased. This further strengthens the claim that our state preparation method is suitable for DPSS states with any dimension, because the high performance observed at low dimensions is predictive of high performance at higher dimensions. 

We have constructed a circuit to efficiently prepare the MPS with a cost that scales logarithmically with both the dimension and desired precision of the prepared state. We have calculated the total number of T gates needed to implement the circuit. Our circuit has reduced non-Clifford gate count compared to prior work and avoids relying on quantum read-only memory, which has a high ancilla qubit overhead. Our circuit is applicable for preparing an MPS approximation of any state with real-valued amplitudes and is useful for applications that do not necessarily involve DPSS states or QPE.

In this work, we have concentrated on providing numerical evidence.
There is still the open question of how to obtain a formal proof that these MPS approximations are accurate for any dimension.
\b[Our interpretation is that the DPSS sequences are well approximated by matrix product states because they converge to the prolate spheroidal window function in the continuum limit, which is a smooth function. Recent theoretical work has shown that target states with amplitudes that sample a smooth function can be well approximated with low bond dimension \cite{bohun2026entanglement}].

Future work could also consider how to reduce the number of qubits used when the dimension is not a power of 2.
That appears to be very difficult, because the unary iteration approach requires accessing all qubits of the control state prior to measurement, as do the shift and inequality test operations that suppress unwanted amplitudes.
An alternative direction for improvement is developing better MPS approximations through optimised padding schemes for non-power-of-2 dimensions.
\section{\b[Code availability]}
\b[
Python code for reproducing the results in all Figures and Tables is available on GitHub \cite{code}.
The Python scripts used for generating Figures 6–8 internally call Mathematica's \texttt{SpheroidalPS} function \cite{Mathematica} to implement the padding technique described in Section \ref{paddingsection}. A Mathematica installation is therefore required for these Figures, though no direct interaction with Mathematica is necessary.
The results of prior work, presented in Table \ref{tab:cost}, were produced with Matlab code provided by the authors of Ref.~\cite{berry2024rapid}.]
\section{Author contributions}
KK performed the numerical work and wrote the manuscript, under the supervision of DWB. Both authors contributed to the analysis and revised the text. LLM tools were used to generate some of the code. All of the code was manually reviewed and verified for correctness by the authors. None of the manuscript text is LLM generated, but LLM suggestions were used to improve structure, grammar, and clarity of presentation.
\section{Acknowledgments}
\b[KK worked on this project with funding from Molecular Quantum Solutions ApS and Sydney Quantum Academy, and would like to thank Mark Nicholas Jones and Kristín Björg Arnardóttir for their feedback on this manuscript.] DWB worked on this project under a sponsored research agreement with Google Quantum AI. DWB is also supported by Australian Research Council Discovery Projects DP210101367 and DP220101602. \b[The authors thank Guang Hao Low for helpful comments regarding the cost of preparing magic states.]
\bibliographystyle{quantum}
\bibliography{bib_quantum}
\appendix
\section{Derivation of bounds}\label{appa}
Since $S(d)\equiv\frac{d}{\pi}\sum_{j,k}\text{sinc}(d(j-k))\ket{j}\bra{k}$ is a positive semi-definite operator, then $\{S(d), \openone-S(d)\}$ is a positive operator valued measure and we can use a well-known inequality relating the fidelity of quantum states to the Bhattacharyya coefficient \cite{watrous2018theory}:
\[
\sqrt{\mathcal{F}}&\leq\sqrt{\text{Tr}[\rho_\text{MPS}S(d)]\text{Tr}[\rho_\text{DPSS}S(d)]}+\sqrt{\text{Tr}[\rho_\text{MPS}(\openone -S(d))]\text{Tr}[\rho_\text{DPSS}(\openone -S(d))]}\nonumber\\&=\sqrt{(1-\delta_\text{MPS})(1-\delta_\text{DPSS})}+\sqrt{\delta_\text{MPS}\delta_\text{DPSS}} \, .
\]
Rearranging this equation yields
\[
\sqrt{\mathcal{F}}-\sqrt{\delta_\text{DPSS}\delta_\text{MPS}}\leq\sqrt{(1-\delta_\text{DPSS})(1-\delta_\text{MPS})} \, .
\]
If $\mathcal{F}>\delta_\text{DPSS}\delta_\text{MPS}$, then both sides are positive. This inequality holds for the regime we are interested in, where we have $\mathcal{F}$ quite close to 1 and $\delta_\text{DPSS}\delta_\text{MPS}$ quite close to zero. Therefore, we can square both sides and then divide by $\delta_\text{DPSS}$ to obtain
\[
\mathcal{F}-2\sqrt{\mathcal{F}\delta_\text{DPSS}\delta_\text{MPS}}+\delta_\text{DPSS}\delta_\text{MPS}&\leq 1-\delta_\text{DPSS}-\delta_\text{MPS}+\delta_\text{DPSS}\delta_\text{MPS}\nn
\mathcal{F}-2\sqrt{\mathcal{F}\delta_\text{DPSS}\delta_\text{MPS}}&\leq 1-\delta_\text{DPSS}-\delta_\text{MPS}\nn
\frac{\delta_\text{MPS}}{\delta_\text{DPSS}}-2\sqrt{\mathcal{F}\frac{\delta_\text{MPS}}{\delta_\text{DPSS}}}+1-\frac{1-\mathcal{F}}{\delta_\text{DPSS}}&\leq 0\label{quadin}.
\]
This is a quadratic inequality with respect to $\sqrt{\delta_\text{MPS}/\delta_\text{DPSS}}$ that has critical numbers $w_\pm$ at
\[
w_\pm=\frac{2\sqrt{\mathcal{F}}\pm\sqrt{4\mathcal{F}-4\left(1-\frac{1-\mathcal{F}}{\delta_\text{DPSS}}\right)}}{2}=\sqrt{\mathcal{F}}\pm\sqrt{\mathcal{F}-1+\frac{1-\mathcal{F}}{\delta_\text{DPSS}}} \, .
\]
The critical numbers are real-valued when $\mathcal{F}-1+(1-\mathcal{F})/\delta_\text{DPSS}\geq 0$. This inequality is equivalent to $(1-\mathcal{F})(1-\delta_\text{DPSS})\delta_\text{DPSS}\geq0$, which is always true because $0\leq\mathcal{F}\leq1$ and $0\leq\delta_\text{DPSS}\leq 1$. Therefore, we have a real-valued upper bound of $\sqrt{\delta_\text{MPS}/\delta_\text{DPSS}}$ at
\[
w_+=\sqrt{\mathcal{F}}+\sqrt{\mathcal{F}-1+\frac{1-\mathcal{F}}{\delta_\text{DPSS}}} \, .
\]
Therefore
\[
\frac{\delta_\text{MPS}-\delta_\text{DPSS}}{\delta_\text{DPSS}}\leq (w_+)^2-1=\left(\sqrt{\mathcal{F}}+\sqrt{\frac{1-\mathcal{F}}{\delta_\text{DPSS}}+\mathcal{F}-1}\right)^2-1.
\]
There is also a lower bound for $\sqrt{\delta_\text{MPS}/\delta_\text{DPSS}}$ at $w_-$, however this only implies a positive lower bound for $(\delta_\text{MPS}-\delta_\text{DPSS})/\delta_\text{DPSS}$ when
\[
w_-&\geq1\nn
\sqrt{F}-\sqrt{\mathcal{F}-1+\frac{1-\mathcal{F}}{\delta_\text{DPSS}}}&\geq 1\nn
1-\sqrt{\mathcal{F}}&\geq\sqrt{\mathcal{F}-1+\frac{1-\mathcal{F}}{\delta_\text{DPSS}}}\nn
1-2\sqrt{\mathcal{F}}+\mathcal{F}&\geq\mathcal{F}-1+\frac{1-\mathcal{F}}{\delta_\text{DPSS}}\nn
\delta_\text{DPSS}&\geq \frac{1+\sqrt{\mathcal{F}}}{2},
\]
which does not occur in the regime we are interested in, where we have $\mathcal{F}$ quite close to 1 and $\delta_\text{DPSS}$ quite close to zero.

\section{\b[Classical computation time]}\label{timeappendix}
\begin{figure}[tbh]
    \centering
    \includegraphics[width=1\linewidth]{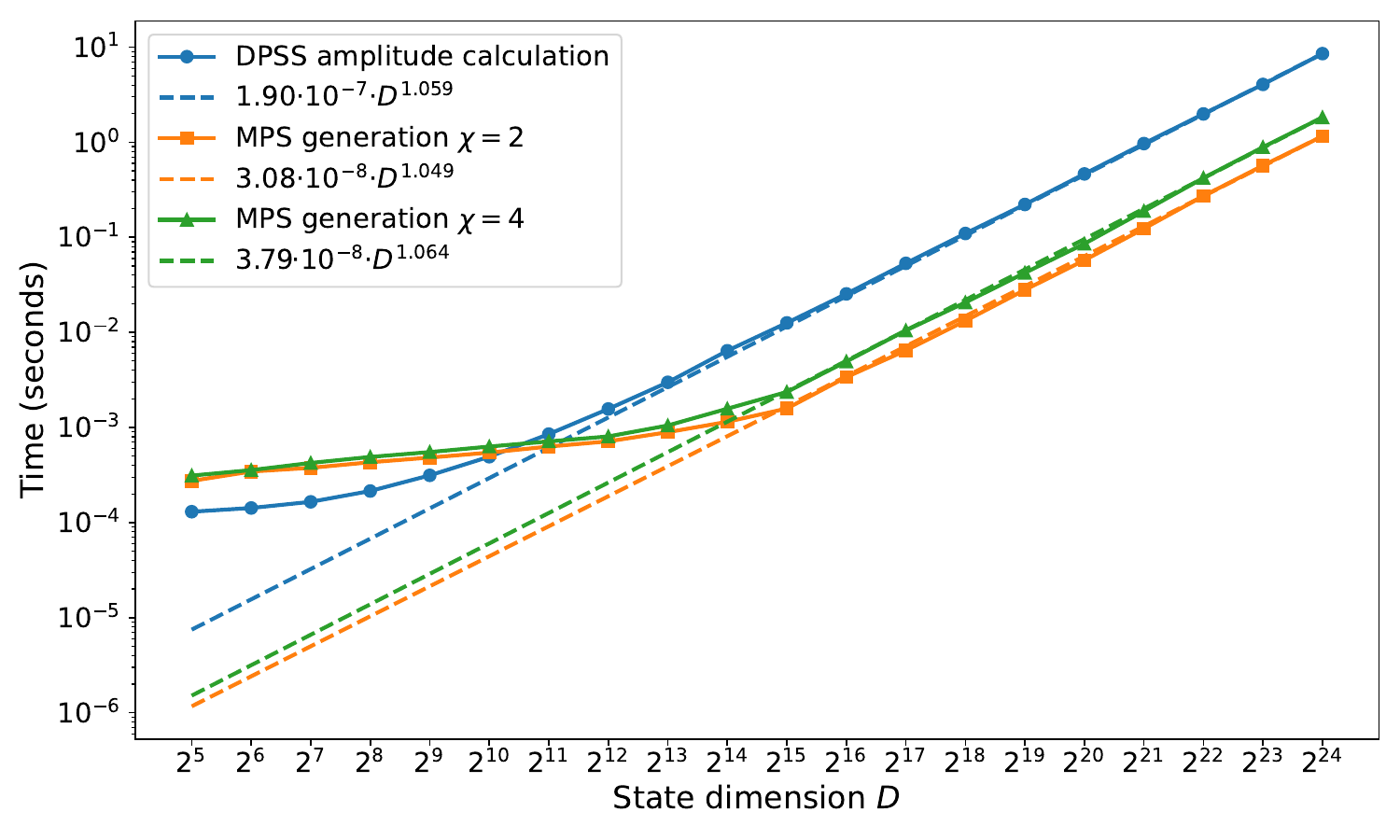}
    \caption{\b[Classical processing time for generating the DPSS amplitudes and MPS operators. The fitted lines are approximately linear in $D$.]}
    \label{fig:time}
\end{figure}

\b[Figure \ref{fig:time} shows the average time used to perform the classical calculations, as a function of state dimension. For the DPSS amplitude calculation, we average the calculation time over three different DPSS states (corresponding to 99\%, 99.9\% and 99.99\% confidence intervals) with 100 trials for each setting. Subsequently, we calculated the isometries for the MPS preparation circuit as described in Section \ref{prep} for bond dimensions $\chi=2$ and $\chi=4$, for a total of 150 states each. We repeated this for power of two state dimensions from $D=2^5$ to $D=2^{24}$.]

\b[The computer used for generating this plot has 256 GB of RAM and a 64-core 2.9 GHz AMD Threadripper 3990X CPU. The DPSS amplitudes are calculated with the \texttt{signal.windows.dpss} function from the SciPy python package and the singular value decompositions are performed with the \texttt{linalg.svd} function from the NumPy Python package. No explicit parallelisation techniques were used for these calculations.]

\section{\b[Example calculation of MPS operators]}\label{appc}
\b[Section \ref{prep} explains how a quantum circuit can be constructed to prepare a matrix product state that approximates a target state. To illustrate this in more detail, we work through an example using the DPSS state with dimension $D=32$, configured to be optimal for a $99\%$ confidence level. Here the bond dimension is $\chi=4$, which is sufficient to prepare the state exactly with no truncation of singular values involved.]

\b[First, we perform the successive applications of reshaping and SVD, as shown in Eq.~\eqref{reshape_and_svd}. We repeat this until we have retrieved all $V^{\dagger[k]}$, which correspond to the rightmost matrices in Eqs.~\eqref{appc1}-\eqref{appc4} below.]
\b[
\[\label{appc1}
\begin{bmatrix}
0.05 & 0.07 \\
0.08 & 0.10 \\
0.12 & 0.13 \\
0.15 & 0.17 \\
0.18 & 0.20 \\
0.21 & 0.22 \\
0.23 & 0.24 \\
0.24 & 0.25 \\
0.25 & 0.24 \\
0.24 & 0.23 \\
0.22 & 0.21 \\
0.20 & 0.18 \\
0.17 & 0.15 \\
0.13 & 0.12 \\
0.10 & 0.08 \\
0.07 & 0.05 \\
\end{bmatrix}
=
\begin{bmatrix}
-0.08 & -0.01 \\
-0.13 & -0.01 \\
-0.17 & -0.01 \\
-0.22 & -0.01 \\
-0.27 & -0.01 \\
-0.31 & -0.01 \\
-0.33 & -0.01 \\
-0.35 & 0.00 \\
-0.35 & 0.00 \\
-0.33 & 0.01 \\
-0.31 & 0.01 \\
-0.27 & 0.01 \\
-0.22 & 0.01 \\
-0.17 & 0.01 \\
-0.13 & 0.01 \\
-0.08 & 0.01 \\
\end{bmatrix}
\begin{bmatrix}
-0.71 & -0.71 \\
0.71 & -0.71 \\
\end{bmatrix}
\]
\[
\resizebox{\linewidth}{!}{$
\begin{bmatrix}
-0.08 & -0.01 & -0.13 & -0.01 \\
-0.17 & -0.01 & -0.22 & -0.01 \\
-0.27 & -0.01 & -0.31 & -0.01 \\
-0.33 & -0.01 & -0.35 & 0.00 \\
-0.35 & 0.00 & -0.33 & 0.01 \\
-0.31 & 0.01 & -0.27 & 0.01 \\
-0.22 & 0.01 & -0.17 & 0.01 \\
-0.13 & 0.01 & -0.08 & 0.01 \\
\end{bmatrix}
=
\begin{bmatrix}
-0.15 & 0.03 & 0.00 & 0.00 \\
-0.28 & 0.04 & 0.00 & 0.00 \\
-0.41 & 0.03 & 0.00 & 0.00 \\
-0.48 & 0.01 & 0.00 & 0.00 \\
-0.48 & -0.01 & 0.00 & 0.00 \\
-0.41 & -0.03 & 0.00 & 0.00 \\
-0.28 & -0.04 & 0.00 & 0.00 \\
-0.15 & -0.03 & 0.00 & 0.00 \\
\end{bmatrix}
\begin{bmatrix}
0.71 & 0.00 & 0.71 & 0.00 \\
0.63 & -0.32 & -0.63 & -0.32 \\
0.00 & -0.71 & 0.00 & 0.71 \\
-0.32 & -0.63 & 0.32 & -0.63 \\
\end{bmatrix}
$}
\]
\[
\resizebox{0.9\linewidth}{!}{$
\begin{aligned}
&\begin{bmatrix}
-0.15 & 0.03 & 0.00 & 0.00 & -0.28 & 0.04 & 0.00 & 0.00 \\
-0.41 & 0.03 & 0.00 & 0.00 & -0.48 & 0.01 & 0.00 & 0.00 \\
-0.48 & -0.01 & 0.00 & 0.00 & -0.41 & -0.03 & 0.00 & 0.00 \\
-0.28 & -0.04 & 0.00 & 0.00 & -0.15 & -0.03 & 0.00 & 0.00 \\
\end{bmatrix}
\\&=
\begin{bmatrix}
-0.30 & -0.11 & 0.01 & 0.00 \\
-0.63 & -0.06 & 0.00 & 0.00 \\
-0.63 & 0.06 & 0.00 & 0.00 \\
-0.30 & 0.11 & 0.01 & 0.00 \\
\end{bmatrix}
\begin{bmatrix}
0.71 & -0.01 & 0.00 & 0.00 & 0.71 & 0.01 & 0.00 & 0.00 \\
-0.62 & -0.34 & 0.00 & 0.00 & 0.62 & -0.34 & 0.00 & 0.00 \\
-0.01 & -0.69 & -0.15 & 0.00 & -0.01 & 0.69 & -0.15 & 0.00 \\
0.24 & -0.45 & -0.49 & -0.05 & -0.24 & -0.45 & 0.49 & -0.05 \\
\end{bmatrix}
\end{aligned}$}
\]
\[\label{appc4}
&\begin{bmatrix}
-0.30 & -0.11 & 0.01 & 0.00 & -0.63 & -0.06 & 0.00 & 0.00 \\
-0.63 & 0.06 & 0.00 & 0.00 & -0.30 & 0.11 & 0.01 & 0.00 \\
\end{bmatrix}
\nn&=
\begin{bmatrix}
-0.66 & -0.26 \\
-0.66 & 0.26 \\
\end{bmatrix}
\begin{bmatrix}
0.71 & 0.04 & 0.00 & 0.00 & 0.71 & -0.04 & 0.00 & 0.00 \\
-0.63 & 0.33 & -0.02 & 0.00 & 0.63 & 0.33 & 0.02 & 0.00 \\
\end{bmatrix}
\]
Finally, we can retrieve $V^{\dagger[1]}$ as the reshaped remainder matrix. The results are
\[
V^{\dagger[5]}&=\begin{bmatrix}
-0.71 & -0.71 \\
0.71 & -0.71 \\
\end{bmatrix}\\
V^{\dagger[4]}&=\begin{bmatrix}
0.71 & 0.00 & 0.71 & 0.00 \\
0.63 & -0.32 & -0.63 & -0.32 \\
0.00 & -0.71 & 0.00 & 0.71 \\
-0.32 & -0.63 & 0.32 & -0.63 \\
\end{bmatrix}\\
V^{\dagger[3]}&=\begin{bmatrix}
0.71 & -0.01 & 0.00 & 0.00 & 0.71 & 0.01 & 0.00 & 0.00 \\
-0.62 & -0.34 & 0.00 & 0.00 & 0.62 & -0.34 & 0.00 & 0.00 \\
-0.01 & -0.69 & -0.15 & 0.00 & -0.01 & 0.69 & -0.15 & 0.00 \\
0.24 & -0.45 & -0.49 & -0.05 & -0.24 & -0.45 & 0.49 & -0.05 \\
\end{bmatrix}\\
V^{\dagger[2]}&=\begin{bmatrix}
0.71 & 0.04 & 0.00 & 0.00 & 0.71 & -0.04 & 0.00 & 0.00 \\
-0.63 & 0.33 & -0.02 & 0.00 & 0.63 & 0.33 & 0.02 & 0.00 \\
\end{bmatrix}\\
V^{\dagger[1]}&=\begin{bmatrix}-0.66& -0.26& -0.66& 0.26\end{bmatrix}
\]
To retrieve the operators $M^{[k]}$ to be used in the quantum circuit, we follow the procedure described in Section \ref{prep}:
\begin{enumerate}
    \item Apply the conjugate transpose to retrieve $V^{[k]}=(V^{\dagger[k]})^\dagger$.
    \item Pad the matrices $V^{[k]}$ so that they are square. Here we will pad them with values denoted as $*$, which indicate that the actual values corresponding to the operator implemented by the quantum circuit do not affect the final result. Instead of padding on one side, pad equally between the existing columns. This ensures the correct ordering of the columns for $M^{[k]}$ as shown in Figure \ref{vandw}.
    \item Reorder the rows by making the last bit of each row the first bit. For example, row $6$ becomes row $3$ because $110\rightarrow011$. This achieves the swapping of $s_k$ with $\alpha_k$ as shown in Figure \ref{vandw}.
\end{enumerate}
The result of performing these steps is shown below:
\[
M^{[5]}&=\begin{bmatrix}
-0.71 & 0.71 \\
-0.71 & -0.71 \\
\end{bmatrix}
\\M^{[4]}&=\begin{bmatrix}
0.71 & 0.63 & 0.00 & -0.32 \\
0.00 & -0.32 & -0.71 & -0.63 \\
0.71 & -0.63 & 0.00 & 0.32 \\
0.00 & -0.32 & 0.71 & -0.63 \\
\end{bmatrix}
\\M^{[3]}&=\begin{bmatrix}
0.71 & * & -0.62 & * & -0.01 & * & 0.24 & * \\
-0.01 & * & -0.34 & * & -0.69 & * & -0.45 & * \\
0.00 & * & 0.00 & * & -0.15 & * & -0.49 & * \\
0.00 & * & 0.00 & * & 0.00 & * & -0.05 & * \\
0.71 & * & 0.62 & * & -0.01 & * & -0.24 & * \\
0.01 & * & -0.34 & * & 0.69 & * & -0.45 & * \\
0.00 & * & 0.00 & * & -0.15 & * & 0.49 & * \\
0.00 & * & 0.00 & * & 0.00 & * & -0.05 & * \\
\end{bmatrix}
\\M^{[2]}&=\begin{bmatrix}
0.71 & * & * & * & -0.63 & * & * & * \\
0.04 & * & * & * & 0.33 & * & * & * \\
0.00 & * & * & * & -0.02 & * & * & * \\
0.00 & * & * & * & 0.00 & * & * & * \\
0.71 & * & * & * & 0.63 & * & * & * \\
-0.04 & * & * & * & 0.33 & * & * & * \\
0.00 & * & * & * & 0.02 & * & * & * \\
0.00 & * & * & * & 0.00 & * & * & * \\
\end{bmatrix}
\\M^{[1]}&=\begin{bmatrix}
-0.66 & * & * & * \\
-0.26 & * & * & * \\
-0.66 & * & * & * \\
0.26 & * & * & * \\
\end{bmatrix}
\]
The original state can then be prepared with the circuit shown in Figure \ref{circuit}, which performs the following operation:
\[
\ps=(\openone_{16}\otimes M^{[5]})(\openone_{8}\otimes M^{[4]})(\openone_{4}\otimes M^{[3]})(\openone_{2}\otimes M^{[2]}\otimes\openone_{2})(M^{[1]}\otimes\openone_8)\ket{0}.
\]
It is straightforward to numerically check that this reproduces the original state (the vector corresponding to the left hand side of Eq.~\eqref{appc1}). Code for performing all of these calculations in available on Github \cite{code}.
]
\end{document}